\documentclass[conference]{IEEEtran}

\usepackage{cite}
\usepackage{amsmath,amssymb,amsfonts}
\usepackage{graphicx}
\usepackage{xcolor}

\usepackage{microtype}
\usepackage{url}
\usepackage{booktabs}
\usepackage{tabularx}
\usepackage{listings}
\usepackage[inline]{enumitem}

\usepackage{placeins}
\usepackage{pdflscape}
\usepackage{subcaption}

\usepackage{hyperref}

\let\code\texttt

\newcommand\pipelinescale{0.9}
\newcommand\plotscale{1.0}
\newcommand\tablescale{1.4} %

\lstdefinelanguage{js}{
    sensitive,
    morekeywords={break,continue,delete,else,for,function,if,in,new,return,this,typeof,var,void,while,with}, %
    morekeywords={await,async,case,catch,class,const,default,do,enum,export,extends,finally,from,implements,import,instanceof,let,static,super,switch,throw,try}, %
    morekeywords={false,null,true,boolean,number,undefined,Array,Boolean,Date,Math,Number,String,Object}, %
    morecomment=[l]{//},
    morecomment=[s]{/*}{*/},
    morestring=[b]",
    morestring=[b]',
    morestring=[b]`,
}

\lstdefinelanguage{json}{
    sensitive,
    morekeywords={true,false},
    morestring=[b]",
}

\lstdefinelanguage{oas}{
    sensitive,
    morekeywords={summary,description,servers,parameters}, %
    morekeywords={openapi,info,paths,components,security,tags}, %
    morekeywords={title,version}, %
    morekeywords={url}, %
    morekeywords={\$ref,get,put,post,delete,patch}, %
    morekeywords={tags,operationId,requestBody,responses,deprecated,security}, %
    morekeywords={\$ref}, %
    morekeywords={name,in,required,deprecated,style,explode,schema}, %
    morekeywords={type,properties,items,enum,format,maximum,minimum}, %
    morekeywords={content,required}, %
    morekeywords={schemas,responses,requestBodies,headers,securitySchemes}, %
    morekeywords={type,name,in,scheme,bearerFormat,flows}, %
    morestring=[b]",
}

\begin{document}

\title{Benchmarking Web API Integration Code Generation
}

\author{
    \IEEEauthorblockN{
        Daniel Maninger\IEEEauthorrefmark{1}\IEEEauthorrefmark{2},
        Leon Chemnitz\IEEEauthorrefmark{4}\IEEEauthorrefmark{1},
        Amir Molzam Sharifloo\IEEEauthorrefmark{1},
        Jannis Brugger\IEEEauthorrefmark{1}\IEEEauthorrefmark{2},
        Mira Mezini\IEEEauthorrefmark{1}\IEEEauthorrefmark{2}\IEEEauthorrefmark{3}
    }
    \IEEEauthorblockA{
        \IEEEauthorrefmark{1}Technische Universität Darmstadt, Germany\\
        \IEEEauthorrefmark{2}Hessian Center for Artificial Intelligence (hessian.AI), Germany\\
        \IEEEauthorrefmark{3}National Research Center for Applied Cybersecurity ATHENE, Germany\\
        \IEEEauthorrefmark{4}Pariton AI, Germany\\
        daniel.maninger@tu-darmstadt.de, leon.chemnitz@pariton.ai, amir.molzam@tu-darmstadt.de,\\
        jannis.brugger@tu-darmstadt.de, mezini@cs.tu-darmstadt.de
    }
}

\maketitle

\begin{abstract}
    API integration is a cornerstone of our digital infrastructure, enabling software systems to connect and interact. However, as shown by many studies, writing or generating correct code to invoke APIs, particularly web APIs, is challenging. Although large language models~(LLMs) have become popular in software development, their effectiveness in automating the generation of web API integration code remains unexplored. In order to address this, we present WAPIIBench, a dataset and evaluation pipeline designed to assess the ability of LLMs to generate web API invocation code. Our experiments with several open-source LLMs reveal that generating API invocations poses a significant challenge, resulting in hallucinated endpoints, incorrect argument usage, and other errors. None of the evaluated open-source models was able to solve more than 40\% of the tasks.

\end{abstract}

\begin{IEEEkeywords}
artificial intelligence, software engineering, large language models, code generation, web APIs, benchmarks
\end{IEEEkeywords}

\section{Introduction}
\label{sec:introduction}

Web APIs provide services and functionality through a standardized, HTTP-based interface. Developers leverage them to rapidly create applications (powered by so-called \emph{API integration code}) that seamlessly integrate with a wide range of online services. The market surrounding API integrations is experiencing explosive growth~\cite{marketsandmarkets_api_2024, postman_state_2024}.

However, writing correct API integration code is a tedious and challenging task. Developers need to understand and correctly utilize a significant amount of information to be able to write an \emph{API invocation} in compliance with its specification~\cite{robillard_what_2009, rauf_systematic_2019}. Moreover, the task of ensuring correct API usage is further complicated by the fact that APIs evolve and change over time.

Large language models~(LLMs) have great potential to boost productivity in software development thanks to their ability to assist developers, e.g., by automatically generating program code from natural language descriptions~\cite{cui_effects_2025}. However, LLMs may also hallucinate and make mistakes, raising concerns about correctness~\cite{dou_whats_2024, tambon_bugs_2025}, security~\cite{pearce_asleep_2022, perry_users_2023}, and general quality~\cite{gitclear_coding_2025} of LLM-generated code.

\paragraph{\textbf{Research gap}}

While the capabilities of LLMs have been investigated for many different software engineering tasks, including library (i.e., non-web) API invocations~\cite{zhuo_pop_2023}, their aptitude for web API integrations appears to be rather unexplored (cf. also related work in Section~\ref{sec:related-work}). The question is: Given the pivotal role of web APIs, can we unlock the potential of LLMs for streamlining and accelerating the development processes of software that involves web API integration? As a first step towards answering this question, we need to understand how well existing LLMs support that task. To this end, we formulate the following research question:

\begin{quote}
    \textbf{RQ:} \textit{How correct is LLM-generated web API invocation code, and what types of errors are commonly present in such code?}
\end{quote}

\paragraph{\textbf{Methodology}}

\begin{figure*}
    \centering
    \includegraphics[alt={Flow diagram visualizing the evaluation pipeline described in the caption.}, width=\pipelinescale\textwidth]{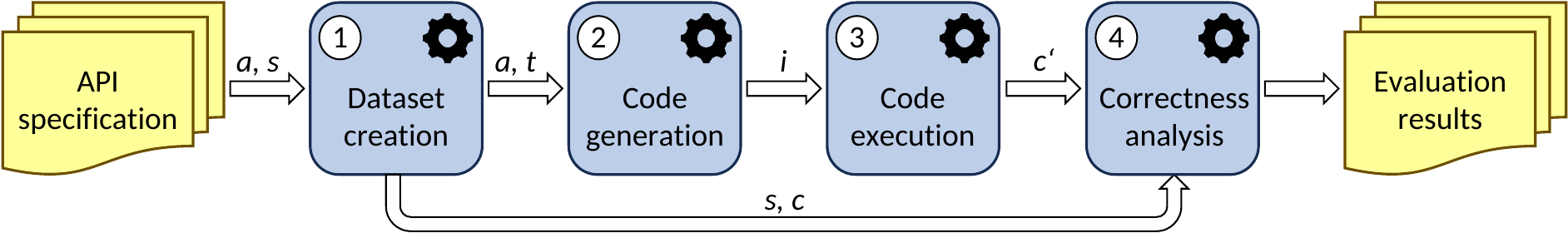}
    \caption{Benchmark design for evaluating the capabilities of LLMs in generating web API invocation code. 1)~Based on an API~$a$ and its specifications~$s$, an API invocation tasks~$t$ and corresponding correct request configurations~$c$ are created. 2)~For each $t$, the LLM under evaluation generates an API invocation~$i$. 3)~$i$ is executed in a controlled environment, yielding a request configuration~$c'$. 4)~$c'$ is compared to $c$ and validated against $s$ to obtain various metrics, shown in Table~\ref{tab:metrics}.}
    \label{fig:eval-pipeline}
\end{figure*}

We investigate the stated research question through the following task: Given a short program context and a code comment that describes a desired operation on a given web API in natural language, the LLM has to complete the code by implementing the operation in compliance with the API's specification.

To arrive at an answer, we create \emph{WAPIIBench}, a benchmark for web API integration tasks. Specifically, we perform the following steps. First, we create a dataset of web API invocation tasks and expected outcomes. Using a powerful state-of-the-art LLM, we generate the initial version of the dataset and subject it to a thorough manual review and correction process. Second, we implement an evaluation pipeline that allows us to safely execute web API invocation code and evaluate the resulting requests to obtain fine-grained correctness metrics. For this purpose, we develop a novel evaluation methodology. Third, we use our dataset and evaluation pipeline to evaluate a comprehensive set of LLMs, including models from the StarCoder~\cite{li_starcoder_2023, lozhkov_starcoder_2024}, DeepSeek-Coder~\cite{guo_deepseek-coder_2024, deepseek-ai_deepseek-coder-v2_2024}, Qwen-Coder~\cite{hui_qwen25-coder_2024}, (Code)~Llama~\cite{roziere_code_2023, dubey_llama_2024}, and GPT families.

\paragraph{\textbf{Scope}}

\begin{lstlisting}[caption={General structure of an API invocation using Axios and JavaScript.\\\phantom{xxxxxx}}, label=lst:invocation-example, language=js, float]
const axios = require('axios');
axios.<method>('http://server.com/path/to/endpoint/arg1', {
    arg2: 'request body parameters'
}, {
    headers: {
        arg3: 'header parameters'
    },
    params: {
        arg4: 'query parameters'
    }
});
\end{lstlisting}

Since web APIs can be used in many ways, we need to be more specific and focus our research specifically on the following setup: We target real-world web APIs that adhere to \emph{OpenAPI}\footnote{\url{https://www.openapis.org/}}, the most widely used specification standard for web APIs~\cite{smartbear_state_2020}. We generate code in \emph{JavaScript}, one of the most popular programming languages~\cite{statista_most_2024}, and use \emph{Axios}\footnote{\url{https://axios-http.com/}}, the leading JavaScript library for invoking web APIs~\cite{devographics_state_2022}, which allows explicit configuration of all components of an HTTP requests.

The general structure of the API invocations we consider is shown in Listing~\ref{lst:invocation-example}. An invocation comprises an HTTP method (GET, POST, etc.), a URL with server address and (potentially parameterized) path to the endpoint, a request body, a request header, and query parameters. Such web API invocations have some unique challenges compared to function calls in a local codebase:
\begin{enumerate*}
    \item Operations are identified by the combination of a method and a long URL string, not just a simple function name;
    \item there is not just one but multiple argument lists and the number of available arguments can be much higher while many arguments have complex, nested data types; and
    \item web API endpoints are documented externally, limiting accessibility for LLMs, and the used specification format is much more complex than, e.g., that of documentation comments used for local functions.
\end{enumerate*}

\paragraph{\textbf{Evaluation results}}

Our study reveals that, without any in-context information about the APIs, most LLMs demonstrate some understanding of the assigned tasks, show familiarity with the general structure of invocations in JavaScript/Axios, and appear to have memorized portions of the APIs' specifications. However, even the best open-source models generate correct invocations only 40\% of the time. In particular, they frequently hallucinate endpoint URLs (up to 39\%) and parameter names (up to 31\%).

\paragraph{\textbf{Contributions}}

In summary, by introducing WAPIIBench, we make the following contributions:

\begin{description}%
    \item[A first-of-its-kind dataset of web API invocation tasks] across four real-world APIs with nearly 400 pairs of tasks and expected outcomes, which can be used for execution-based correctness analysis. \textrightarrow~Section~\ref{subsec:dataset}

    \item[An open-source evaluation pipeline] for automatically and safely evaluating the performance of code generation models on web API invocation tasks using fine-grained metrics. \textrightarrow~Section~\ref{subsec:generation} -- Section~\ref{subsec:analysis}

    \item[Novel empirical insights] on the (in)ability of LLMs to generate correct API invocation code. \textrightarrow~Section~\ref{sec:evaluation}
\end{description}

WAPIIBench is available on GitHub\footnote{\url{https://github.com/stg-tud/WAPIIBench}}.
In addition, we provide all model-generated codes in our artifact\footnote{\url{https://doi.org/10.5281/zenodo.17607587}}. Additional details and comprehensive results are provided in an appendix at the end of this paper. %

\section{Benchmark Design}
\label{sec:benchmark}

WAPIIBench's evaluation pipeline, depicted in Figure~\ref{fig:eval-pipeline}, consists of four stages:
\begin{enumerate*}
    \item \emph{dataset creation},
    \item \emph{code generation},
    \item \emph{code execution}, and
    \item \emph{correctness analysis}.
\end{enumerate*}
They are detailed in the following subsections.

\subsection{Dataset Creation}
\label{subsec:dataset}

We conceive of our dataset as a collection of triples~$(a, t, c)$, where $a$ is the name of the targeted API, $t$ is a task, given as an unambiguous natural language description of a request to $a$, and $c$ is a so-called request \emph{configuration}, which captures all properties an API request must have in order to solve $t$---it is the key to measuring functional correctness. Specifically, $c$ contains the expected URL, HTTP method, and arguments in the request body, header, and query locations. In this paper, we use the term \emph{endpoint} to refer to the combination of URL and (HTTP) method.

To our knowledge, no dataset with the described structure exists. To create a new one, two general approaches exist: sourcing data from the real world or synthesizing it. We first explored the former option by searching public GitHub repositories for suitable JavaScript/Axios code\footnote{Using BigQuery (\url{https://cloud.google.com/bigquery/}) for this purpose.}. We dropped this option and resorted to the latter for two reasons. First, we found that real-world code rarely includes sufficient information (e.g., comments or in-code documentation) to serve as an unambiguous task description~$t$. Second, the resulting dataset would have been incomplete, not covering all endpoints of a given API.

Instead, we opted for a synthetic approach, using an LLM~$\operatorname{Gen}$ and  API specifications~$s$ to generate task--configuration pairs for our dataset: $\operatorname{Gen}(a, s) = (t, c)$. To ensure our dataset represents real-world scenarios, we chose APIs of popular services across different domains---Asana, Google Calendar, Google Sheets, and Slack---which are specified following OpenAPI, the most commonly used industry standard~\cite{smartbear_state_2020}. As $\operatorname{Gen}$, we used Gemini~1.5~Pro\footnote{\url{https://blog.google/technology/ai/google-gemini-next-generation-model-february-2024/}} for its very large context window (2M tokens), given that OpenAPI specifications can be very long (e.g., Slack's contains over 20,000 lines of text). Together with the full specification, we provided the model with very detailed (but mostly non-API-specific) instructions on how to construct $t$ and $c$ in a zero-shot prompt to avoid pitfalls of few-shot prompting~\cite{reynolds_prompt_2021} and increase generalizability. The prompt is provided in the appendix. In total, we synthesized 395 samples---one for each endpoint of the selected APIs. A detailed breakdown is shown in Table~\ref{tab:endpoints}.

To ensure the validity of the AI-generated data, we curated all samples with automated and manual checks. The automated checks caught any inconsistencies between the generated configurations~$c$ and the specifications~$s$. The 9 samples that failed these checks were manually inspected and corrected. Afterward, we manually checked all 395 samples to catch remaining issues that cannot be checked automatically, especially underspecification and ambiguities in the tasks~$t$, in order to ensure that there is one unique $c$ for each $t$. Concretely, we checked
\begin{enumerate*}
    \item that $t$ is well-formulated and solvable;
    \item that $c$ actually solves $t$; and
    \item that every argument in $t$ is reflected in $c$ and vice versa.
\end{enumerate*}
58 samples had one or multiple issues and were fixed by adjusting $t$ and/or $c$.

Listing~\ref{lst:code-example} shows a simplified example from our dataset, with the API~$a$ being \textit{Google Calendar}. The task~$t$ is given as a comment, followed by the implementation the model has to generate. Listing~\ref{lst:config-example} shows the corresponding configuration~$c$, which contains key--value pairs specifying the correct HTTP method, URL, and parameters---in this case, there are only parameters in the request body (``\code{data}'').

\begin{table}
    \centering
    \caption{Number of Endpoints per HTTP Method for Each API (Each Endpoint Corresponds to One Sample in the Dataset)}
    \label{tab:endpoints}
    \begin{tabularx}{\linewidth}{Xrrrrrr}
        \toprule
        API & Total & POST & GET & PUT & \scalebox{0.88}[1.0]{DELETE} & \scalebox{0.88}[1.0]{PATCH} \\
        \midrule
        Asana & 167 & 61 & 79 & 14 & 13 & 0 \\
        Google Calendar & 37 & 14 & 11 & 4 & 4 & 4 \\
        Google Sheets & 17 & 12 & 4 & 1 & 0 & 0 \\
        Slack Web & 174 & 95 & 79 & 0 & 0 & 0 \\
        \midrule
        Total & 395 & 182 & 173 & 19 & 17 & 4 \\
        \bottomrule
    \end{tabularx}
\end{table}

\subsection{Code Generation}
\label{subsec:generation}

We take each pair of API~$a$ and task~$t$ from our dataset and let the model under evaluation~$\operatorname{M}$ generate an API invocation~$i$: $\operatorname{M}(a, t) = i$. The prompt instructs $M$ to solve $t$ by completing the starter code that is given below it. We specify $a$ in order to disambiguate $t$ as otherwise a common task like ``create a new user \dots'' could refer to many different APIs. The prompt is provided in the appendix. We also include some clarifications in the prompt, meant to avoid instances of misalignment and recurring error patterns we observed in preliminary experiments. The starter code always includes $t$ as a line comment (e.g., \code{// Create a new secondary calendar \dots}), besides the required imports.

We conduct our experiments with two different starter code variants---we call them \emph{setups}---that serve complementary purposes. \emph{Full completion} includes the beginning of API invocation (\code{axios.}). By fixing the code up to this position, we ensure that an Axios call is actually generated and can be evaluated. In this setup, we can assess the ability of the model to identify the correct endpoint for $t$ and to solve the task as a whole. \emph{Argument completion} additionally includes the correct method and URL (e.g., \code{axios.post('https://\allowbreak{}www.googleapis.com/\allowbreak{}calendar/\allowbreak{}v3/\allowbreak{}calendars',}). In this setup, we can evaluate whether the model is able to use a given endpoint in compliance with $a$. Also, this setup resembles real-world scenarios inside an IDE where the user has already implemented a part of the solution and wants it to be completed by the AI.

Note that the code generation task described here is considerably different from the dataset generation task from Section~\ref{subsec:dataset}. In the latter, Gemini had access to the full API specification and very detailed instructions, making this a task about processing in-context information. The models under evaluation, on the other hand, have to rely solely on memorized knowledge about the APIs, as we want to find out how well they can recall these details from their training data. This is a relevant concern as context length is limited, and it is often not desirable to fill the context with too much information. For instance, Slack's specification has around 126K tokens, exceeding the context length of most models we evaluated.

\begin{lstlisting}[caption={Simplified task from our dataset with correct implementation.\\\phantom{xxxxxx}}, label=lst:code-example, language=js, float]
// Create a new secondary calendar named "Example Calendar" with time zone "America/Los_Angeles".
const axios = require('axios');
axios.post(
      'https://www.googleapis.com/calendar/v3/calendars', {
    summary: 'Example Calendar',
    timeZone: 'America/Los_Angeles'
});
\end{lstlisting}

\begin{lstlisting}[caption={Request configuration corresponding to the task in Listing~\ref{lst:code-example}.\\\phantom{xxxxxx}}, label=lst:config-example, language=json, float]
{
  "method": "post",
  "url":
      "https://www.googleapis.com/calendar/v3/calendars",
  "data": {
    "summary": "Example Calendar",
    "timeZone": "America/Los_Angeles"
  }
}
\end{lstlisting}

\subsection{Code Execution}
\label{subsec:execution}

Our approach to evaluating LLMs' capabilities in generating API invocation code is based on \emph{functional correctness} rather than syntactic similarity to a reference implementation, as the latter is prone to misjudgments~\cite{evtikhiev_out_2023}. To measure functional correctness, we need to execute the generated code and assess its outcome, analogous to unit testing~\cite{chen_evaluating_2021}.

When executing $i$, we need to address challenges unique to our setting. First, direct execution of generated code is unsafe as it could send arbitrary requests to external servers. Second, we need to capture and evaluate the configuration~$c'$ of the API request implemented by $i$, which is a \emph{side effect} and not the return value of $i$. Another common challenge when evaluating code generation is \emph{excess code}~\cite{guo_when_2024}, i.e., code produced when the model terminates too late. 

To address these challenges, we implement a controlled execution environment~$\operatorname{Mock}$, which intercepts requests before transmission and then serializes and returns their configurations: $\operatorname{Mock}(i) = c'$. We address the \emph{excess code} issue in $\operatorname{Mock}$ by heuristically truncating the generated code after the API invocation, preventing the execution of unnecessary, incomplete, broken, or harmful code.

\subsection{Correctness Analysis}
\label{subsec:analysis}

\begin{table}
    \centering
    \caption{Evaluation Metrics for Correctness and Specification-Compliance of API Invocations}
    \label{tab:metrics}
    \begin{tabularx}{\linewidth}{p{0.32\linewidth}X}
        \toprule
        Metric & Description \\
        \midrule
        Correct implementations & Generated executable code matches the ground-truth configuration \\
        Illegal implementations & Generated code contains at least one violation of the API specification \\
        Correct URLs & Generated URL matches the ground-truth URL \\
        Illegal URLs & Generated URL is not defined in the API specification \\
        Correct methods & Generated HTTP method matches the ground-truth HTTP method \\
        Illegal methods & Generated HTTP method is not defined for the generated URL in the API specification \\
        Mean argument precision & Probability that the generated arguments are correct \\
        Mean argument recall & Probability that the correct arguments are generated \\
        Illegal arguments & Generated arguments are not permitted for the generated API endpoint \\
        Mean argument value conditional accuracy & Probability that an argument value is correct if the argument name is correct \\
        \bottomrule
    \end{tabularx}
\end{table}

We compare the configuration~$c'$, captured in the previous step, with the ground-truth $c$ from our dataset: $c' \overset{?}{=} c$. However, as web API requests contain multiple components---URL, method, and parameters in different locations (cf. Listing~\ref{lst:invocation-example})---only looking for exact matches is insufficient. Instead, we compare the configurations in an element-wise manner as well as in aggregated forms to facilitate a fine-grained evaluation. Additionally, we validate $c'$ against the API specification~$s$ to distinguish between endpoints and arguments that are undesired but specification-compliant and ones that are actually illegal. The key metrics we calculate are explained in Table~\ref{tab:metrics} and additional ones are provided in the appendix. Note that some metrics apply only to one of the setups (full/argument completion), as mentioned in Section~\ref{subsec:generation}.

\section{Evaluation}
\label{sec:evaluation}

Using our dataset and evaluation pipeline, we study the performance of 21 state-of-the-art open-source LLMs on web API invocation code generation. We primarily evaluated models that have undergone fine-tuning on code data---CodeT5+~\cite{wang_codet5_2023}~(16B), StarCoder~\cite{li_starcoder_2023}~(15.5B), StarCoder2~\cite{lozhkov_starcoder_2024}~(3B, 7B, 15B), DeepSeek-Coder~\cite{guo_deepseek-coder_2024}~(1.3B, 6.7B, 7B, 33B), DeepSeek-Coder-V2~\cite{deepseek-ai_deepseek-coder-v2_2024}~(16B), Qwen2.5-Coder~\cite{hui_qwen25-coder_2024}~(0.5B, 1.5B, 3B, 7B, 14B, 32B), and Code~Llama~\cite{roziere_code_2023}~(7B, 13B, 70B)---and additionally Llama~3.1~\cite{dubey_llama_2024}~(8B, 70B). These models were selected due to their frequent appearance in coding leaderboards\footnote{E.g., \url{https://evalplus.github.io/leaderboard.html}}. To obtain an upper bound on performance, we also evaluated OpenAI's commercial GPT-4o\footnote{\url{https://openai.com/index/hello-gpt-4o/}} and GPT-4o mini\footnote{\url{https://openai.com/index/gpt-4o-mini-advancing-cost-efficient-intelligence/}} models. For a lack of space and for more visual clarity, we show only a subset of these models in our plots and tables, but our following analysis considers all of them. Comprehensive result tables are provided in the appendix.

Throughout this paper, we use the notation \emph{(t)} to indicate that a metric is calculated based on the \emph{total} amount of samples in the dataset and \emph{(e)} to indicate that it is calculated based on only the samples that resulted in \emph{executable} code\footnote{To be exact, the calculation for (t) and (e) values is identical except for a filtering step to remove non-executable samples before calculating the latter.}. We do this because our evaluation method can only assess the correctness of individual request components if the generated code can be executed, and would otherwise classify them as incorrect by default. The distinction between (t) and (e) allows us to differentiate between a model's ability to generate \emph{complete} and \emph{executable} code~(t), and a model's ability to generate code that \emph{complies with the specification} and \emph{solves the given task}, provided that it is executable~(e). For illustration why this distinction is necessary, consider a model~$\operatorname{M}$ that predicts the correct endpoint 100\% of the time but introduces syntax errors in the request arguments 50\% of the time. When looking at (t) metrics only, we would wrongly conclude that $\operatorname{M}$ predicts endpoints correctly only 50\% of the time.

\subsection{Results}
\label{subsec:results}

\begin{figure}
    \centering
    \includegraphics[alt={Parallel coordinates plot visualizing the experimental results from Table~\ref{tab:res-multi-inv}.}, width=\plotscale\columnwidth]{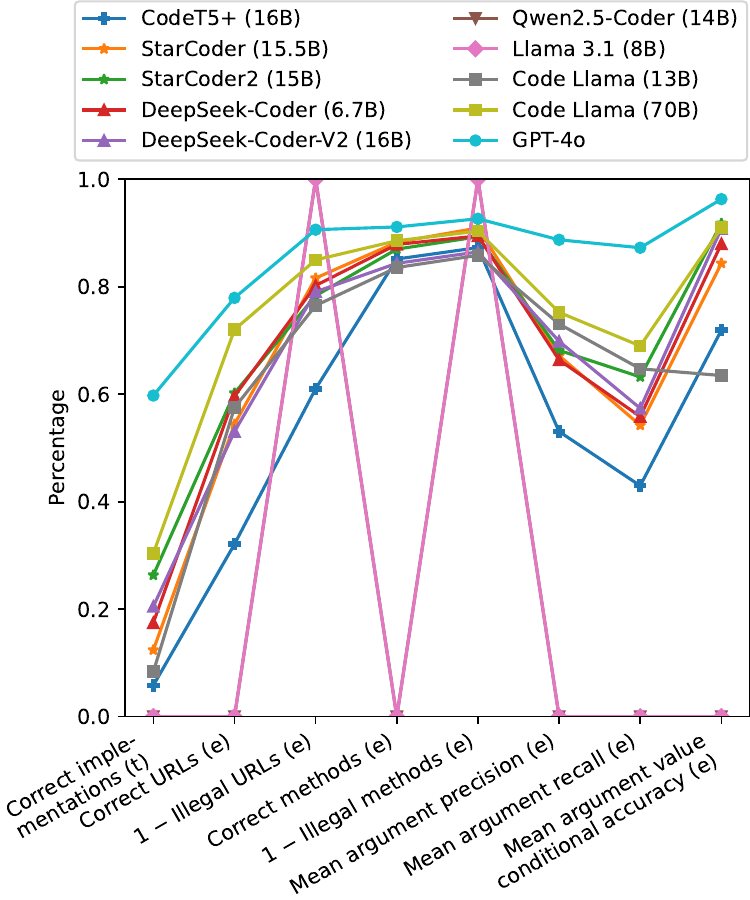}
    \includegraphics[alt={Parallel coordinates plot visualizing the experimental results from Table~\ref{tab:res-multi-end}}, width=\plotscale\columnwidth]{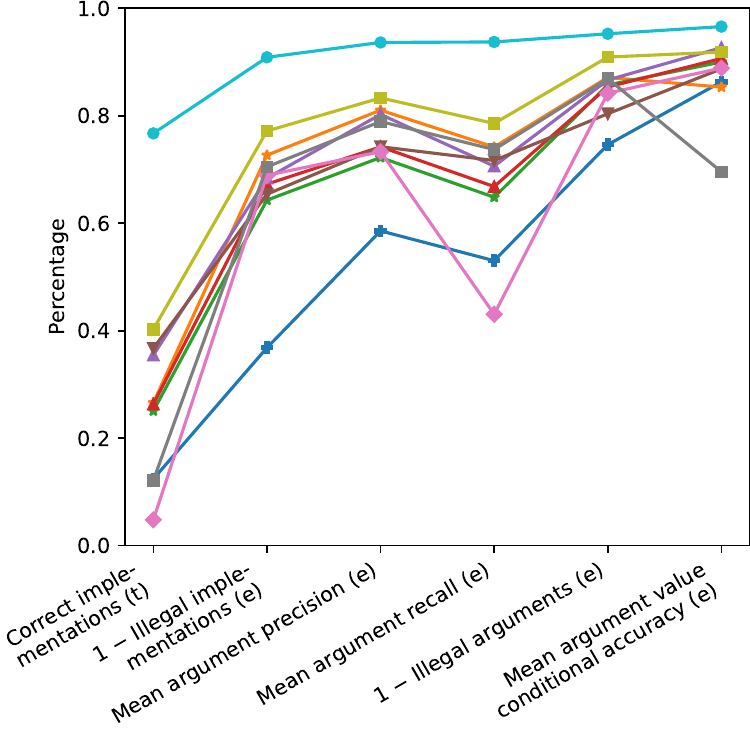}
    \caption{Performance comparison between selected models for \emph{full completion} (top) and \emph{argument completion} (bottom).}
    \label{fig:res-multi}
\end{figure}

Figure~\ref{fig:res-multi} shows key performance metrics (explained in Table~\ref{tab:metrics}) for one representative\footnote{With a size as close as possible to 15B for a fair comparison.} of each model family plus the best-performing open-source model for full and argument completion, respectively. The corresponding numbers are provided in Tables~\ref{tab:res-multi-inv} and~\ref{tab:res-multi-end}. We used greedy decoding for all experiments.

\paragraph{\textbf{Full completion}}

Across all evaluated models, the overall achieved correctness---correct implementations~(t)---ranges from 0\% for the Qwen2.5-Coder and Llama~3.1 models up to 60\% for GPT-4o, which also dominates on all other metrics. The best open-source model is Code~Llama~(70B) at 30\% followed by StarCoder2 at 26\%. In all cases with 0\% correctness, (almost) none of the generated implementations are executable---since also all other metrics for these models become zero, we omit reporting their numbers in the following. For the remaining models, the executability is 94\% and higher.

Most open-source models correctly generate over 80\% of HTTP methods~(e). The rate of correct URLs~(e) varies strongly between 32\% and 72\%, but 14\% to 39\% of URLs are illegal~(e). All models have higher precision~(e) than recall~(e) for arguments, ranging from 51\% to 75\% and from 33\% to 69\%, respectively. If an argument name was predicted correctly, the argument value was correct 53\%--84\% of the time~(e). For all of these metrics, Code~Llama~(70B) is the strongest open-source model, even surpassing GPT-4o~mini. DeepSeek-Coder~(7B) and CodeT5+~(16B) perform the worst across individual metrics.

\paragraph{\textbf{Argument completion}}

Compared to full completion, almost all performance metrics are better for argument completion. The rate of correct implementations~(t) ranges from 8\% for DeepSeek-Coder~(7B) to 40\% for Code~Llama~(70B), which is again the strongest open-source model. The best performing model overall is again GPT-4o at 77\% correctness. The models that struggled to generate any executable code before now achieve similar executability rates as the other models. Qwen2.5-Coder~(14B) and (32B) are now even among the better-performing models.

Overall, for all metrics except correct implementations~(t), we observe a smaller variance between the open-source models and an even clearer performance gap between them and GPT-4o. Since in the argument completion setup, the correct endpoint is provided to the models, we can now evaluate their ability to generate argument lists without violating the API specification. We observe that 6\% to 31\% of generated arguments are illegal~(e).

\subsection{Discussion}
\label{subsec:discussion}

In summary, all tested open-source LLMs struggle with generating correct API invocations. While they often generate partially correct implementations---indicating that the models have had training exposure to API specifications or examples from which they memorized usage patterns of the respective APIs---the models fail to combine individual patterns to fully correct API invocations. The full completion setup shows that the models often do not find the correct endpoint to use. But even when the correct endpoint is provided in the argument completion setup, the models still often fail to stay consistent with the given endpoint and its defined parameters.

The models perform comparatively well in predicting the HTTP method, which is expected since there is only a small set of available methods, with GET and POST being by far the most common ones (cf.~Table~\ref{tab:endpoints}). In contrast, URLs and arguments are challenging to predict due to their unrestricted nature. The argument precision may be inflated to some extent by the \code{Authorization} header argument, which is required by most endpoints and therefore easy to predict. The relatively high argument value conditional accuracy indicates that identifying the correct argument is harder for the models than assigning the correct value to it once it is found.

An analysis of the generated implementations shows that those models that got nearly zero implementations correct in the full completion setup failed to understand the task given to them, resulting in incomplete or syntactically broken and thus non-executable code. Specifically, Qwen2.5-Coder mostly refused to continue the provided starter code, while Llama~3.1 skipped over the method part of the API invocation. It may be possible to solve this issue through prompt engineering, but we preferred to use a single prompt consistently for all models.

An additional finding is that larger models are not always better. In multiple model families (DeepSeek-Coder, Qwen2.5-Coder, Code~Llama), we observe that the medium-sized variant performs worse than both the smaller and larger variants. Lastly, breaking down the evaluation results by API (cf. appendix), we see that each model is particularly good or bad at different APIs, which likely results from the prevalence of the respective APIs in the model's training data. Overall, most models perform best on the Google Calendar API.

\begin{quote}
    \textbf{RQ:} \textit{How correct is LLM-generated API invocation code, and what types of errors are commonly present in such code?}

    \textbf{Answer:} While LLMs are able to generate partially correct API invocations, the overall result is incorrect most of the time. The types of errors commonly present are manifold and range from selecting the wrong endpoint to leaving out required arguments or hallucinating entirely illegal arguments.
\end{quote}

\begin{table*}
\centering
\caption{Evaluation Results for \emph{Full Completion}}
\label{tab:res-multi-inv}
\begin{tabular}{lrrrrrrrrrr}
\makebox[20pt][l]{\rotatebox{20}{}} & \makebox[20pt][l]{\rotatebox{20}{CodeT5+ (16B)}} & \makebox[20pt][l]{\rotatebox{20}{StarCoder (15.5B)}} & \makebox[20pt][l]{\rotatebox{20}{StarCoder2 (15B)}} & \makebox[20pt][l]{\rotatebox{20}{DeepSeek-Coder (6.7B)}} & \makebox[20pt][l]{\rotatebox{20}{DeepSeek-Coder-V2 (16B)}} & \makebox[20pt][l]{\rotatebox{20}{Qwen2.5-Coder (14B)}} & \makebox[20pt][l]{\rotatebox{20}{Code Llama (13B)}} & \makebox[20pt][l]{\rotatebox{20}{Code Llama (70B)}} & \makebox[20pt][l]{\rotatebox{20}{Llama 3.1 (8B)}} & \makebox[20pt][l]{\rotatebox{20}{GPT-4o}} \\
\midrule
Correct implementations (t) & \phantom{00}0.06 & \phantom{00}0.12 & \phantom{00}0.26 & \phantom{00}0.18 & \phantom{00}0.21 & \phantom{00}0.00 & \phantom{00}0.08 & \phantom{00}0.30 & \phantom{00}0.00 & \phantom{00}0.60 \\
Correct URLs (e) & \phantom{00}0.32 & \phantom{00}0.54 & \phantom{00}0.60 & \phantom{00}0.60 & \phantom{00}0.53 & \phantom{00}0.00 & \phantom{00}0.58 & \phantom{00}0.72 & \phantom{00}0.00 & \phantom{00}0.78 \\
Illegal URLs (e) & \phantom{00}0.39 & \phantom{00}0.18 & \phantom{00}0.22 & \phantom{00}0.20 & \phantom{00}0.21 & \phantom{00}0.00 & \phantom{00}0.24 & \phantom{00}0.15 & \phantom{00}0.00 & \phantom{00}0.09 \\
Correct methods (e) & \phantom{00}0.85 & \phantom{00}0.88 & \phantom{00}0.87 & \phantom{00}0.88 & \phantom{00}0.84 & \phantom{00}0.00 & \phantom{00}0.84 & \phantom{00}0.89 & \phantom{00}0.00 & \phantom{00}0.91 \\
Illegal methods (e) & \phantom{00}0.13 & \phantom{00}0.09 & \phantom{00}0.11 & \phantom{00}0.11 & \phantom{00}0.14 & \phantom{00}0.00 & \phantom{00}0.14 & \phantom{00}0.10 & \phantom{00}0.00 & \phantom{00}0.07 \\
Mean argument precision (e) & \phantom{00}0.52 & \phantom{00}0.65 & \phantom{00}0.68 & \phantom{00}0.65 & \phantom{00}0.67 & \phantom{00}0.00 & \phantom{00}0.72 & \phantom{00}0.75 & \phantom{00}0.00 & \phantom{00}0.89 \\
Mean argument recall (e) & \phantom{00}0.43 & \phantom{00}0.54 & \phantom{00}0.63 & \phantom{00}0.56 & \phantom{00}0.57 & \phantom{00}0.00 & \phantom{00}0.65 & \phantom{00}0.69 & \phantom{00}0.00 & \phantom{00}0.87 \\
Mean argument value conditional accuracy (e) & \phantom{00}0.53 & \phantom{00}0.68 & \phantom{00}0.79 & \phantom{00}0.72 & \phantom{00}0.76 & \phantom{00}0.00 & \phantom{00}0.57 & \phantom{00}0.84 & \phantom{00}0.00 & \phantom{00}0.95 \\
\bottomrule
\end{tabular}
\end{table*}

\begin{table*}
\centering
\caption{Evaluation Results for \emph{Argument Completion}}
\label{tab:res-multi-end}
\begin{tabular}{lrrrrrrrrrr}
\makebox[20pt][l]{\rotatebox{20}{}} & \makebox[20pt][l]{\rotatebox{20}{CodeT5+ (16B)}} & \makebox[20pt][l]{\rotatebox{20}{StarCoder (15.5B)}} & \makebox[20pt][l]{\rotatebox{20}{StarCoder2 (15B)}} & \makebox[20pt][l]{\rotatebox{20}{DeepSeek-Coder (6.7B)}} & \makebox[20pt][l]{\rotatebox{20}{DeepSeek-Coder-V2 (16B)}} & \makebox[20pt][l]{\rotatebox{20}{Qwen2.5-Coder (14B)}} & \makebox[20pt][l]{\rotatebox{20}{Code Llama (13B)}} & \makebox[20pt][l]{\rotatebox{20}{Code Llama (70B)}} & \makebox[20pt][l]{\rotatebox{20}{Llama 3.1 (8B)}} & \makebox[20pt][l]{\rotatebox{20}{GPT-4o}} \\
\midrule
Correct implementations (t) & \phantom{00}0.12 & \phantom{00}0.27 & \phantom{00}0.25 & \phantom{00}0.26 & \phantom{00}0.35 & \phantom{00}0.37 & \phantom{00}0.12 & \phantom{00}0.40 & \phantom{00}0.05 & \phantom{00}0.77 \\
Illegal implementations (e) & \phantom{00}0.63 & \phantom{00}0.27 & \phantom{00}0.36 & \phantom{00}0.33 & \phantom{00}0.32 & \phantom{00}0.35 & \phantom{00}0.30 & \phantom{00}0.23 & \phantom{00}0.31 & \phantom{00}0.09 \\
Mean argument precision (e) & \phantom{00}0.58 & \phantom{00}0.81 & \phantom{00}0.71 & \phantom{00}0.74 & \phantom{00}0.77 & \phantom{00}0.74 & \phantom{00}0.79 & \phantom{00}0.83 & \phantom{00}0.65 & \phantom{00}0.94 \\
Mean argument recall (e) & \phantom{00}0.53 & \phantom{00}0.74 & \phantom{00}0.65 & \phantom{00}0.67 & \phantom{00}0.71 & \phantom{00}0.72 & \phantom{00}0.74 & \phantom{00}0.79 & \phantom{00}0.43 & \phantom{00}0.94 \\
Illegal arguments (e) & \phantom{00}0.25 & \phantom{00}0.13 & \phantom{00}0.14 & \phantom{00}0.15 & \phantom{00}0.13 & \phantom{00}0.20 & \phantom{00}0.13 & \phantom{00}0.09 & \phantom{00}0.16 & \phantom{00}0.05 \\
Mean argument value conditional accuracy (e) & \phantom{00}0.71 & \phantom{00}0.83 & \phantom{00}0.83 & \phantom{00}0.83 & \phantom{00}0.86 & \phantom{00}0.78 & \phantom{00}0.65 & \phantom{00}0.89 & \phantom{00}0.67 & \phantom{00}0.95 \\
\bottomrule
\end{tabular}
\end{table*}

\subsection{Limitations and Threats to Validity}
\label{subsec:limitations}

1) Real-world API integration code is diverse and complex. To be able to evaluate API invocations automatically, we had to apply some restrictions and simplifications when creating the tasks in our dataset, as described in Section~\ref{subsec:dataset}. For instance, there is limited program context before the API invocation. Additionally, we evaluate only the correctness of the outgoing request, not how the incoming response is handled. These limitations do, however, not interfere with our investigation of LLM's ability to recall information about correct API usage---if this information is memorized, they should be able to recall it in our evaluation as well as real-world settings.

2) The quality of the dataset and evaluation depends on the quality of the API specifications. We encountered and manually fixed several errors in the real-world specifications. Moreover, the specifications sometimes contain usage constraints that are only explained in a parameter's free-form textual description and thus not captured by our automated correctness analysis.

3) Despite thorough vetting, our dataset might contain faulty samples that could introduce noise into the evaluation results. We also noticed that the synthetic tasks produced by Gemini~1.5~Pro tend to use optional parameters rather sparingly and often use placeholder values instead of realistic example values, limiting the transferability of findings from our benchmark to real-world API invocation tasks. Moreover, the results for some models may not generalize to other APIs, given the limited number of APIs included in our dataset. As models can be very sensitive to prompt variations, the performance of each model can almost certainly be optimized by customized prompt engineering.

\section{Related Work}
\label{sec:related-work}

\paragraph{\textbf{Differentiation of API Invocation Settings}}
\label{subsec:rw-invocations}

Since APIs are such a universal concept, it is important to distinguish between different settings in which APIs are being used and understand their different characteristics when discussing related work. We divide APIs roughly into the following categories:
\begin{enumerate*}
    \item \emph{general web APIs} (this work),
    \item \emph{domain-specific web APIs},
    \item \emph{SDK-wrapped web APIs},
    \item \emph{local function APIs}, and
    \item \emph{tool APIs for AI agents}.
\end{enumerate*}

1) Our work targets general web API invocations, which involve explicitly configuring all components of an HTTP request, i.e., an endpoint URL, HTTP method, and multiple request argument lists with oftentimes complex nested data types. To our knowledge, we are the first to explore and to propose a method for evaluating the capabilities of LLMs in generating code with such specific and challenging characteristics.

2) Su et~al.~\cite{su_building_2017, su_natural_2018} and Hosseini et~al.~\cite{hosseini_compositional_2021} investigated natural language interfaces to domain-specific web APIs following the Open Data Protocol\footnote{\url{https://docs.oasis-open.org/odata/odata/v4.0/odata-v4.0-part1-protocol.html}}~(OData). Their work has a narrow focus on data-centric applications (mostly within the Microsoft ecosystem) and an SQL-like syntax. In contrast, our benchmark targets OpenAPI, a very general and flexible API standard, making the task more open-ended and challenging. Also, instead of ad-hoc API queries by a user, we strive for production-ready API integration code implemented in JavaScript.

3) Works like Dialog2API~\cite{shu_dialog2api_2022} and CloudAPIBench~\cite{jain_mitigating_2025} focused on SDK-wrapped web API invocations (e.g., for AWS), treating web APIs like local libraries. Both approaches do not generalize due to SDK heterogeneity. In contrast, our benchmark relies on a unified RESTful interface for \emph{any} web API. This eliminates SDK dependency, enabling direct assessment of models' core API comprehension rather than SDK-specific knowledge. By standardizing interactions across diverse APIs, we remove evaluation biases and isolate model capabilities from implementation quirks of individual SDKs.

4) Local function APIs are the most common kind of API in typical codebases and studied in many works, e.g.,~\cite{zan_private-library-oriented_2025, eghbali_-hallucinator_2024, dutta_applying_2024}. Invoking a function requires specifying the function name and one argument list. In contrast, general web API invocations are subject to more complex syntax and semantics than such local function calls, as already mentioned in Section~\ref{sec:introduction} (\textit{Scope}).

5) There is a large body of work concerned with enabling LLM-based AI agents to use \emph{tools} to interact with their environment~\cite{yehudai_survey_2025}. These tools range from simple utilities like calculators~\cite{schick_toolformer_2023} to web services that enable AI agents to perform tasks such as weather queries or restaurant booking~\cite{song_restgpt_2023, li_api-bank_2023, qin_toolllm_2024}. At first sight, this body of work seems related to research on automatic generation of integration code, as tools typically expose a function-like interface. However, both input and output modalities differ between these settings, as well as their non-functional requirements, so benchmarks are not directly transferable to other kinds of API invocations.

\paragraph{\textbf{API Invocation Datasets}}
\label{subsec:rw-datasets}

Since functional abstractions are essential in programming, virtually every code dataset includes (function) API invocations. Code datasets involving \emph{web} APIs can be found in the fields of natural language interfaces to web APIs~\cite{su_building_2017, hosseini_compositional_2021}, SDK-based API invocations~\cite{shu_dialog2api_2022, jain_mitigating_2025}, and tool-using AI agents~\cite{patil_gorilla_2024, song_restgpt_2023, li_api-bank_2023, qin_toolllm_2024, yan_bcfl_2024}. We are not aware of datasets specifically about generating API integration code to invoke REST APIs like we do.

\paragraph{\textbf{Evaluation Methods for Web API Invocations}}
\label{subsec:rw-eval}

The correctness of AI-generated code is traditionally measured either using exact matches or syntax-based similarity, which both correlate weakly with actual correctness~\cite{evtikhiev_out_2023}, or using functional testing~\cite{chen_evaluating_2021}, which is not directly transferable to web API invocations, as discussed in Section~\ref{subsec:execution}. This is why we developed a novel approach described in Section~\ref{sec:benchmark}. Evaluation methods in related works on API invocations mostly rely on variations of the aforementioned standard approaches. Some works use humans~\cite{ding_toolcoder_2025} or LLMs~\cite{qin_toolllm_2024} as judges, and others execute the API invocation and verify the state of the environment~\cite{shu_dialog2api_2022, yan_bcfl_2024}. We are not aware of an evaluation method able to provide results as fine-grained as ours.

\paragraph{\textbf{Studies on Memorization and Hallucination in Code LLMs}}
\label{subsec:rw-insights}

LLMs display a remarkable capacity to \emph{memorize} information from their training data---which can be beneficial when it comes to generating factual responses, but also a problem when it comes to leaking private information from the training data. Studies have shown that LLMs memorize, among other things, method definitions and method calls~\cite{yang_unveiling_2024}, but also that they struggle to suggest the correct APIs~\cite{zhuo_pop_2023}. These findings are in line with our observation that LLMs seem to know many endpoint URLs and argument names, but are often unable to assemble them to a specification-compliant API invocation that solves the task in question.

If LLMs are unable to recall memorized facts, they tend to \emph{hallucinate} information that is not grounded in reality. This is a common error of LLMs, and studies have shown that it also exists in code generation, where, among other things, models may hallucinate function names and arguments~\cite{liu_exploring_2024, dou_whats_2024, tambon_bugs_2025}. Our study confirms that these findings also hold for web API endpoints and their arguments, where a high amount of hallucinations can be observed. 

\paragraph{\textbf{Symbolic Approaches to Detecting API Misuse}}
\label{subsec:rw-symbolic}

Correct usage of APIs is challenging also for humans, as indicated by several studies on API usability issues~\cite{robillard_what_2009, rauf_systematic_2019} and real-world API misuses~\cite{bae_safewapi_2014, li_large-scale_2021, wang_apicad_2023}. Several (static analysis) methods have been proposed for detecting (and mitigating) API misuses, e.g.,~\cite{bae_safewapi_2014, binder_jguard_2022, wang_apicad_2023}. They are, however, primarily designed to aid humans, not LLMs. Using static analyses for evaluation purposes is generally conceivable, but they can only detect whether an API invocation is legal or illegal, not whether it also solves the given task.

\section{Conclusion and Future Work}
\label{sec:conclusion}

In this paper, we examined the challenges faced by large language models when generating reliable web API invocation code. We introduced WAPIIBench, a novel dataset and evaluation pipeline to systematically assess LLM performance in this domain. Our experiments with open-source LLMs revealed significant limitations: Only 30\% (full completion) and 40\% (argument completion) of the produced samples were completely right. Endpoints and arguments are often hallucinated due to insufficient recall of knowledge on API usage.

Overall, our work underscores the need for integrating quality assurance techniques into AI-driven code generation workflows to ensure correctness and reliability. Our benchmark lays the foundation for future work to investigate the impact of different approaches---such as retrieval-augmented generation~\cite{lewis_retrieval-augmented_2020}, fine-tuning, reasoning, feedback loops, and constrained decoding~\cite{hokamp_lexically_2017, deutsch_general-purpose_2019}---on the correctness and specification-compliance of generated web API invocation code. While our evaluation pipeline is specialized on JavaScript and Axios, our dataset and methodology is generic and can be transferred to other languages and libraries.

\section*{Acknowledgments}

This work was funded by the Hessian Ministry of Higher Education, Research, Science and the Arts within the cluster project \textit{The Third Wave of Artificial Intelligence} (3AI), by the National Research Center for Applied Cybersecurity ATHENE within the project \textit{Foundational Models for Secure Software Development}, and by the LOEWE initiative (Hesse, Germany) [LOEWE/4a//519/05/00.002(0013)/95].

\bibliographystyle{IEEEtran}
\bibliography{bibliography}

\clearpage

\appendix

\subsection{Data Availability}
\label{app:artifact}

WAPIIBench is available on GitHub\footnote{\url{https://github.com/stg-tud/WAPIIBench}}. In addition, we provide all model-generated codes in our artifact\footnote{\url{https://doi.org/10.5281/zenodo.17607587}}.

\subsection{Background}
\label{app:background}

\begin{figure*}
    \centering
    \begin{subfigure}{\textwidth}
        \begin{lstlisting}[language=oas]
openapi: 3.0.0
servers:
  - url: https://www.googleapis.com/calendar/v3
info:
  title: Calendar API
  description: Manipulates events and other calendar data.
paths:
  /calendars:
    post:
      description: Creates a secondary calendar.
      parameters:
        - name: prettyPrint
          description: Returns response with indentations and line breaks.
          in: query
          required: false
          schema:
            type: boolean
      requestBody:
        content:
          application/json:
            schema:
              properties:
                summary:
                  description: Title of the calendar.
                  type: string
                timeZone:
                  description: The time zone of the calendar. (Formatted as an IANA Time Zone Database name, e.g. "Europe/Zurich".) Optional.
                  type: string
              type: object
      security:
        - Oauth2:
            - https://www.googleapis.com/auth/calendar
        \end{lstlisting}
        \caption{OpenAPI specification (YAML)}
        \label{lst:spec}
    \end{subfigure}
    \par
    \begin{subfigure}{0.48\textwidth}
        \begin{lstlisting}[language=js]
// Create a secondary calendar with summary "Example Calendar" and time zone "America/Los_Angeles". Pretty print the response.
const axios = require('axios');

axios.post('https://www.googleapis.com/calendar/v3/calendars', {
  summary: 'Example Calendar',
  timeZone: 'America/Los_Angeles',
}, {
  headers: {
    Authorization: 'Bearer <access_token>'
  },
  params: {
    prettyPrint: true,
  }
}).then(response => {
  console.log('Calendar created', response.data);
});
        \end{lstlisting}
        \caption{API invocation (JavaScript)}
        \label{lst:code}
    \end{subfigure}
    \hfill
    \begin{subfigure}{0.48\textwidth}
        \begin{lstlisting}[language=json]


{
  "headers": {
    "Accept": "application/json, text/plain, */*",
    "Content-Type": "application/json",
    "Authorization": "Bearer <access_token>"
  },
  "params": {
    "prettyPrint": true
  },
  "method": "post",
  "url": "https://www.googleapis.com/calendar/v3/calendars",
  "data": {
    "summary": "Example Calendar",
    "timeZone": "America/Los_Angeles"
  }
}
        \end{lstlisting}
        \caption{Configuration object (JSON)}
        \label{lst:config}
    \end{subfigure}
    \caption{Example of a web API and its usage. \textbf{Top:} Excerpt from the Google Calendar OpenAPI specification. \textbf{Left:} JavaScript code to send a request to this API using the Axios library. \textbf{Right:} Configuration object that describes the request sent. For our evaluation, we pair the task description (comment in the JavaScript code) with the configuration to create an input--output sample.}
    \label{fig:spec-code-config}
\end{figure*}

OpenAPI provides a structured, human- and machine-readable way to document an API's endpoints, requests, response formats, authentication methods, and other details. For this paper, we define the term \emph{endpoint} as a unique combination of path and HTTP method. Specification files are written in JSON or YAML. Listing~\ref{lst:spec} shows an excerpt of the OpenAPI specification of the Google Calendar API\footnote{Adapted from \url{https://api.apis.guru/v2/specs/googleapis.com/calendar/v3/openapi.yaml}; the original comprises 3190 lines.}. Besides some meta information, it contains a list of paths (e.g., \code{/calendars}) and, for each path, a list of supported HTTP methods (e.g., \code{post}). Each endpoint has properties for documentation purposes and a list of named parameters (\code{prettyPrint}). Parameters can be passed in different locations, indicated by the property \code{in}: \code{path} parameters are inserted directly into of the URL, \code{query} parameters are appended to the URL as key--value pairs, and \code{header} parameters become part of the HTTP request header. Additionally, methods like POST use the \code{requestBody} to transfer data to the server (e.g., \code{summary}, \code{timeZone}). We consider this data as another kind of parameter. Parameters have a \code{schema} that specifies their data type and constraints on permissible values. The \code{security} property determines available authentication schemes, such as OAuth~2.0.

Listing~\ref{lst:code} shows how a request to the endpoint from the specification in Listing~\ref{lst:spec} could be implemented in JavaScript using the Axios library. The method called on the \code{axios} object determines the HTTP method, and the first argument determines the server URL and path. This is followed by one object containing the request body and another object containing header and query parameters (\code{headers}, \code{params}). Query parameters are automatically serialized and appended to the URL by Axios. If we execute the code in Listing~\ref{lst:code}, a configuration object as shown in Listing~\ref{lst:config} is created and a request, configured accordingly, is sent to the given URL. The configuration contains all arguments explicitly given in the code, as well as some implicit parameters, such as \code{Accept} (for the expected response media type) and \code{Content-Type} (for the request body media type). The representation of requests as configuration objects is key to our evaluation method described in the paper.

\subsection{Technologies}

We implemented WAPIIBench using the following technologies:
\begin{itemize}[noitemsep] %
	\item \href{https://www.openapis.org/}{OpenAPI}
	\item \href{https://github.com/manchenkoff/openapi3-parser}{OpenAPI 3 parser}
	\item \href{https://axios-http.com/}{Axios}
	\item \href{https://github.com/ctimmerm/axios-mock-adapter}{axios-mock-adapter}
	\item \href{https://huggingface.co/docs/transformers/index}{Hugging Face Transformers}
\end{itemize}

\subsection{Models}
\label{app:models}

These are the exact names of the models we evaluated:
\begin{itemize}[noitemsep]
	\item \href{https://huggingface.co/bigcode/starcoderbase}{\code{bigcode/starcoderbase}}
	\item \href{https://huggingface.co/bigcode/starcoder2-3b}{\code{bigcode/starcoder2-3b}}
    \item \href{https://huggingface.co/bigcode/starcoder2-7b}{\code{bigcode/starcoder2-7b}}
    \item \href{https://huggingface.co/bigcode/starcoder2-15b}{\code{bigcode/starcoder2-15b}}
	\item \href{https://huggingface.co/deepseek-ai/deepseek-coder-1.3b-base}{\code{deepseek-ai/deepseek-coder-1.3b-base}}
    \item \href{https://huggingface.co/deepseek-ai/deepseek-coder-6.7b-base}{\code{deepseek-ai/deepseek-coder-6.7b-base}}
	\item \href{https://huggingface.co/deepseek-ai/deepseek-coder-7b-base-v1.5}{\code{deepseek-ai/deepseek-coder-7b-base-v1.5}}
	\item \href{https://huggingface.co/deepseek-ai/deepseek-coder-33b-base}{\code{deepseek-ai/deepseek-coder-33b-base}}
	\item \href{https://huggingface.co/deepseek-ai/DeepSeek-Coder-V2-Lite-Base}{\code{deepseek-ai/DeepSeek-Coder-V2-Lite-Base}}
    \item \href{https://ai.google.dev/gemini-api/docs/models#gemini-1.5-pro}{\code{google/gemini-pro-1.5}}
	\item \href{https://huggingface.co/codellama/CodeLlama-7b-hf}{\code{meta-llama/CodeLlama-7b-hf}}
	\item \href{https://huggingface.co/codellama/CodeLlama-13b-hf}{\code{meta-llama/CodeLlama-13b-hf}}
	\item \href{https://huggingface.co/codellama/CodeLlama-70b-hf}{\code{meta-llama/CodeLlama-70b-hf}}
	\item \href{https://huggingface.co/meta-llama/Llama-3.1-8B}{\code{meta-llama/Llama-3.1-8B}}
	\item \href{https://huggingface.co/meta-llama/Llama-3.1-70B}{\code{meta-llama/Llama-3.1-70B}}
	\item \href{https://platform.openai.com/docs/models/gpt-4o}{\code{openai/gpt-4o}}
	\item \href{https://platform.openai.com/docs/models/gpt-4o-mini}{\code{openai/gpt-4o-mini}}
	\item \href{https://huggingface.co/Qwen/Qwen2.5-Coder-0.5B}{\code{Qwen/Qwen2.5-Coder-0.5B}}
	\item \href{https://huggingface.co/Qwen/Qwen2.5-Coder-1.5B}{\code{Qwen/Qwen2.5-Coder-1.5B}}
	\item \href{https://huggingface.co/Qwen/Qwen2.5-Coder-3B}{\code{Qwen/Qwen2.5-Coder-3B}}
	\item \href{https://huggingface.co/Qwen/Qwen2.5-Coder-7B}{\code{Qwen/Qwen2.5-Coder-7B}}
	\item \href{https://huggingface.co/Qwen/Qwen2.5-Coder-14B}{\code{Qwen/Qwen2.5-Coder-14B}}
	\item \href{https://huggingface.co/Qwen/Qwen2.5-Coder-32B}{\code{Qwen/Qwen2.5-Coder-32B}}
	\item \href{https://huggingface.co/Salesforce/codet5p-16b}{\code{Salesforce/codet5p-16b}}
\end{itemize}

As can be seen, we use only base (i.e., non-instruction-tuned) models. While instruction-tuned models tend to perform better on coding tasks\footnote{Cf., e.g., the EvalPlus leaderboard (\url{https://evalplus.github.io/leaderboard.html})}, we considered them to be inappropriate for our setting, which is based on \emph{code completion}. Performing code completion on instruction-tuned models leads to rather unnatural results, as the models often do not directly return the completion and instead start with some response (``Here is your completed code ...'') followed by a code snipped that may contain
\begin{enumerate*}
    \item only the completion,
    \item the starter code followed by the completion, or
    \item an arbitrarily modified version of the starter code and corresponding completion.    
\end{enumerate*}
These factors make the model output very hard to parse reliably. Therefore, we decided to focus our evaluation on base models.

The following hyperparameters were used:
\begin{itemize}[noitemsep] %
    \item floating point precision = 16 bit
    \item \# beams = 1
    \item temperature = 0.0
\end{itemize}

\subsection{Prompts}
\label{app:prompts}

\begin{lstlisting}[caption={Prompt for generating the test data for a given API. Words in curly braces are placeholders.}, label=lst:dataset-prompt, frame=none, float]
Consider this OpenAPI specification:

```yaml
{spec}
```

I want you to generate test data for this API. The data should be in JSON format and look like this:

```json
{
  "samples": [
    {
      "task": "...",
      "config": {
        "url": "...",
        "method": "...",
        "headers": { ... },
        "params": { ... },
        "data": { ... }
      }
    },
    ...
  ]
}
```

* `samples` is an array containing all the test cases.
* `task` is a natural language description of a specific task that can be solved by sending a request to the given API. All information required to unambiguously identify and implement the corresponding API request must be contained in the task description. Therefore, state all expected argument values explicitly. The only exception is authentication keys and tokens, which must not be specified here (but you may specify the authentication method to be used if multiple ones are available).
* `config` is an object containing the expected configuration of the described request. If a property of `config` is empty, it can be omitted.
* `url` is the full URL of the endpoint (server URL + path) and may include path parameters but no query parameters.
* `method` is the HTTP method used. It can be one of `get`, `put`, `post`, `delete`, and `patch`.
* `headers` is an object containing all expected header arguments. Note that the property `"Accept": "application/json, text/plain, */*"` is always present and if `data` is sent in the request body, `"Content-Type": "mime/type"` is present as well (substitute "mime/type" for the respective media type).
* `params` is an object containing all expected query arguments.
* `data` is an object containing all data from the request body. It is only used for the methods put, post, delete, and patch.

Remember that for authentication, an additional argument might be required. Take a look at an endpoint's `security` property and the `securitySchemes` section in the specification, to find out if and how to authenticate. I assume you know how the usual authentication schemes `apiKey`, `http`, and `oauth2` work. Use `<key>` as a placeholder for API keys (e.g., `"name": "<key>"`) and `<token>` as a placeholder for authorization tokens (e.g., `"Authorization": "Bearer <token>"`).
{api_specific_instructions}
Now, please give me a JSON object that matches my description and that contains diverse examples of requests to this API.
{path_selection}
\end{lstlisting}

\begin{lstlisting}[caption={Prompt for generating API invocations given a task description. Words in curly braces are placeholders.}, label=lst:evaluation-prompt, frame=none, float]
You are an AI programming assistant that helps users write API requests. You are given a comment that describes what the user wants to achieve and are supposed to implement it using the Axios library in JavaScript. For this, write a single call to Axios (with syntax `axios.method(url[, config])`) that does exactly what was described in the comment.

* Make sure to include all parameters in `config` that are required to solve the given task but don't include any unnecessary parameters.
* Insert all values directly into the place where they belong, rather than using intermediate variables.
* If the API requires some form of authentication, use `<key>` as a placeholder for API keys or `<token>` as a placeholder for authorization tokens, respectively.
* If a request body requires a media type other than `text/json`, explicitly set the `Content-Type` header to the respective type, and Axios will automatically serialize the request body accordingly.

Your next task is about the {api} API. Complete the following code snippet{extra_instructions}:

```javascript
// {task}
const axios = require('axios');

axios.{method}('{url}',
\end{lstlisting}

Listing~\ref{lst:dataset-prompt} shows the prompt used for creating the dataset with Gemini~1.5~pro. To avoid incomplete responses when generating the dataset based on OpenAPI specifications with more than 100 endpoints (Asana and Slack), we prompted it multiple times, asking for different subsets of endpoints. The Slack API required additional instructions to account for changes in the API that are not reflected in the OpenAPI specification\footnote{\url{https://api.slack.com/changelog/2017-10-keeping-up-with-the-jsons}}. Listing~\ref{lst:evaluation-prompt} shows the prompt used in our evaluation pipeline when generating API invocation code.

In both cases, we decided against few-shot prompting to increase generalizability and to avoid associated pitfalls. Few-shot prompting techniques mainly help the model to understand the task it is supposed to solve and the desired response format. This is not the issue in the type of task we investigate. Rather, the main problem is that models select the wrong API endpoints and/or pass the wrong arguments to the endpoint. On the other hand, using few-shot prompts introduces new challenges, such as a high sensitivity to the examples provided while reducing the generalizability of the prompt and not guaranteeing inherently superior performance to that of zero-shot prompting~\cite{reynolds_prompt_2021}.

Further optimization of model performance via prompt engineering would require either revealing information about the API’s endpoints and parameters---undermining the purpose of our evaluation of the models’ memorized knowledge---or tailoring prompts to each individual model (e.g., adjusting wording, formatting, etc.). However, this approach relies on trial-and-error and introduces high uncertainty and variance.

\subsection{Extended Results}
\label{app:results}

The full set of metrics calculated by our pipeline is explained in Table~\ref{tab:metrics-extended}. Tables~\ref{tab:res-multi-ud-inv} and~\ref{tab:res-multi-ud-end} provide a comprehensive summary of these metrics for all evaluated models\footnote{While we also evaluated Gemini~1.5~Pro on WAPIIBench, we excluded it from all discussion---since it was the model that generated the dataset, its evaluation results might be skewed.} (cf. Appendix~\ref{app:models}). Additionally, the results are broken down by API exemplarily for StarCoder2 and GPT-4o in Tables~\ref{tab:res-sc-ud-inv} to~\ref{tab:res-4o-ud-end}. The best values are in bold. Note that some metrics can only be calculated either for \emph{full completion} or for \emph{argument completion} and are therefore not shown in all tables. Values marked with \emph{(t)} are ratios relative to the 395 \emph{total samples}, while values marked with \emph{(e)} are relative to the subset of \emph{executable codes}, which varies from experiment to experiment. The complete raw data these results are based on can be found in our artifact (cf. Appendix~\ref{app:artifact}).

\begin{table*}
    \centering
    \caption{Complete Evaluation Metrics for Correctness and Specification-Compliance of API Invocations}
    \label{tab:metrics-extended}
    \begin{tabular}{ll}
        \toprule
        Metric & Description \\
        \midrule
        Executable codes & Generated code is complete and contains no syntax or runtime error \\
        Correct codes & Generated executable code matches the ground-truth configuration \\
        Illegal codes & Generated code contains at least one violation of the API specification \\
        Correct URLs & Generated URL matches the ground-truth URL \\
        Illegal URLs & Generated URL is not defined in the API specification \\
        Correct methods & Generated HTTP method matches the ground-truth HTTP method \\
        Illegal methods & Generated HTTP method is not defined for the generated URL in the API specification \\
        Correct argument names & Generated arguments are correct \\
        Correct argument values & Generated argument values are correct \\
        Missing arguments & Expected arguments are not generated \\
        Unexpected arguments & Generated arguments are not expected \\
        Unnecessary arguments & Redundant arguments are generated for an API endpoint \\
        Illegal arguments & Generated arguments are not permitted for the generated API endpoint \\
        Mean argument precision & Probability that the generated arguments are correct \\
        Mean argument recall & Probability that the correct arguments are generated \\
        Mean argument Jaccard index & Overlap between generated and correct arguments \\
        Mean argument value conditional accuracy & Probability that an argument value is correct if the argument name is correct \\
        Total errors & Any type of error prevented execution \\
        Incomplete codes & Generated code did not contain a complete API invocation \\
        Runtime errors & Generated code produced an error when trying to execute it \\
        \bottomrule
    \end{tabular}
\end{table*}

\begin{landscape}
\begin{table}[p]
\centering
\caption{Complete Evaluation Results for \emph{Full Completion}}
\label{tab:res-multi-ud-inv}
\makebox[1\textwidth][c]{
\resizebox{\tablescale\textwidth}{!}{
\begin{tabular}{lrrrrrrrrrrrrrrrrrrrrrrrr}
\makebox[20pt][l]{\rotatebox{45}{}} & \makebox[20pt][l]{\rotatebox{45}{CodeT5+ (16B)}} & \makebox[20pt][l]{\rotatebox{45}{StarCoder (15.5B)}} & \makebox[20pt][l]{\rotatebox{45}{StarCoder2 (3B)}} & \makebox[20pt][l]{\rotatebox{45}{StarCoder2 (7B)}} & \makebox[20pt][l]{\rotatebox{45}{StarCoder2 (15B)}} & \makebox[20pt][l]{\rotatebox{45}{DeepSeek-Coder (1.3B)}} & \makebox[20pt][l]{\rotatebox{45}{DeepSeek-Coder (6.7B)}} & \makebox[20pt][l]{\rotatebox{45}{DeepSeek-Coder (7B)}} & \makebox[20pt][l]{\rotatebox{45}{DeepSeek-Coder (33B)}} & \makebox[20pt][l]{\rotatebox{45}{DeepSeek-Coder-V2 (16B)}} & \makebox[20pt][l]{\rotatebox{45}{Qwen2.5-Coder (0.5B)}} & \makebox[20pt][l]{\rotatebox{45}{Qwen2.5-Coder (1.5B)}} & \makebox[20pt][l]{\rotatebox{45}{Qwen2.5-Coder (3B)}} & \makebox[20pt][l]{\rotatebox{45}{Qwen2.5-Coder (7B)}} & \makebox[20pt][l]{\rotatebox{45}{Qwen2.5-Coder (14B)}} & \makebox[20pt][l]{\rotatebox{45}{Qwen2.5-Coder (32B)}} & \makebox[20pt][l]{\rotatebox{45}{Code Llama (7B)}} & \makebox[20pt][l]{\rotatebox{45}{Code Llama (13B)}} & \makebox[20pt][l]{\rotatebox{45}{Code Llama (70B)}} & \makebox[20pt][l]{\rotatebox{45}{Llama 3.1 (8B)}} & \makebox[20pt][l]{\rotatebox{45}{Llama 3.1 (70B)}} & \makebox[20pt][l]{\rotatebox{45}{Gemini 1.5 Pro}} & \makebox[20pt][l]{\rotatebox{45}{GPT-4o mini}} & \makebox[20pt][l]{\rotatebox{45}{GPT-4o}} \\
\midrule
Executable implementations (t) & \phantom{00}0.95 & \phantom{00}0.94 & \phantom{00}0.97 & \phantom{00}0.95 & \phantom{00}0.99 & \phantom{00}0.96 & \phantom{00}0.96 & \phantom{00}0.57 & \phantom{00}0.98 & \phantom{00}0.95 & \phantom{00}0.01 & \phantom{00}0.00 & \phantom{00}0.00 & \phantom{00}0.00 & \phantom{00}0.00 & \phantom{00}0.00 & \phantom{00}0.99 & \phantom{00}0.89 & \phantom{00}0.99 & \phantom{00}0.00 & \phantom{00}0.00 & \textbf{1.00} & \textbf{1.00} & \textbf{1.00} \\
Correct implementations (t) & \phantom{00}0.06 & \phantom{00}0.12 & \phantom{00}0.11 & \phantom{00}0.10 & \phantom{00}0.26 & \phantom{00}0.08 & \phantom{00}0.18 & \phantom{00}0.04 & \phantom{00}0.17 & \phantom{00}0.21 & \phantom{00}0.00 & \phantom{00}0.00 & \phantom{00}0.00 & \phantom{00}0.00 & \phantom{00}0.00 & \phantom{00}0.00 & \phantom{00}0.12 & \phantom{00}0.08 & \phantom{00}0.30 & \phantom{00}0.00 & \phantom{00}0.00 & \phantom{00}0.45 & \phantom{00}0.39 & \textbf{\phantom{00}0.60} \\
Correct implementations (e) & \phantom{00}0.06 & \phantom{00}0.13 & \phantom{00}0.12 & \phantom{00}0.11 & \phantom{00}0.27 & \phantom{00}0.08 & \phantom{00}0.18 & \phantom{00}0.06 & \phantom{00}0.17 & \phantom{00}0.21 & \phantom{00}0.00 & \phantom{00}0.00 & \phantom{00}0.00 & \phantom{00}0.00 & \phantom{00}0.00 & \phantom{00}0.00 & \phantom{00}0.12 & \phantom{00}0.09 & \phantom{00}0.31 & \phantom{00}0.00 & \phantom{00}0.00 & \phantom{00}0.45 & \phantom{00}0.39 & \textbf{\phantom{00}0.60} \\
Correct URLs (t) & \phantom{00}0.31 & \phantom{00}0.51 & \phantom{00}0.40 & \phantom{00}0.45 & \phantom{00}0.60 & \phantom{00}0.33 & \phantom{00}0.57 & \phantom{00}0.35 & \phantom{00}0.67 & \phantom{00}0.51 & \phantom{00}0.01 & \phantom{00}0.00 & \phantom{00}0.00 & \phantom{00}0.00 & \phantom{00}0.00 & \phantom{00}0.00 & \phantom{00}0.48 & \phantom{00}0.51 & \phantom{00}0.72 & \phantom{00}0.00 & \phantom{00}0.00 & \phantom{00}0.72 & \phantom{00}0.57 & \textbf{\phantom{00}0.78} \\
Correct URLs (e) & \phantom{00}0.32 & \phantom{00}0.54 & \phantom{00}0.41 & \phantom{00}0.48 & \phantom{00}0.60 & \phantom{00}0.35 & \phantom{00}0.60 & \phantom{00}0.62 & \phantom{00}0.68 & \phantom{00}0.53 & \phantom{00}0.67 & \phantom{00}0.00 & \phantom{00}0.00 & \phantom{00}0.00 & \phantom{00}0.00 & \phantom{00}0.00 & \phantom{00}0.48 & \phantom{00}0.58 & \phantom{00}0.72 & \phantom{00}0.00 & \phantom{00}0.00 & \phantom{00}0.72 & \phantom{00}0.57 & \textbf{\phantom{00}0.78} \\
Illegal URLs (t) & \phantom{00}0.37 & \phantom{00}0.17 & \phantom{00}0.22 & \phantom{00}0.27 & \phantom{00}0.22 & \phantom{00}0.32 & \phantom{00}0.19 & \phantom{00}0.10 & \phantom{00}0.13 & \phantom{00}0.20 & \textbf{\phantom{00}0.00} & \textbf{\phantom{00}0.00} & \textbf{\phantom{00}0.00} & \textbf{\phantom{00}0.00} & \textbf{\phantom{00}0.00} & \textbf{\phantom{00}0.00} & \phantom{00}0.28 & \phantom{00}0.21 & \phantom{00}0.15 & \textbf{\phantom{00}0.00} & \textbf{\phantom{00}0.00} & \phantom{00}0.19 & \phantom{00}0.23 & \phantom{00}0.09 \\
Illegal URLs (e) & \phantom{00}0.39 & \phantom{00}0.18 & \phantom{00}0.23 & \phantom{00}0.29 & \phantom{00}0.22 & \phantom{00}0.34 & \phantom{00}0.20 & \phantom{00}0.17 & \phantom{00}0.14 & \phantom{00}0.21 & \textbf{\phantom{00}0.00} & \textbf{\phantom{00}0.00} & \textbf{\phantom{00}0.00} & \textbf{\phantom{00}0.00} & \textbf{\phantom{00}0.00} & \textbf{\phantom{00}0.00} & \phantom{00}0.28 & \phantom{00}0.24 & \phantom{00}0.15 & \textbf{\phantom{00}0.00} & \textbf{\phantom{00}0.00} & \phantom{00}0.20 & \phantom{00}0.23 & \phantom{00}0.09 \\
Correct methods (t) & \phantom{00}0.81 & \phantom{00}0.83 & \phantom{00}0.78 & \phantom{00}0.88 & \phantom{00}0.86 & \phantom{00}0.79 & \phantom{00}0.84 & \phantom{00}0.53 & \phantom{00}0.86 & \phantom{00}0.81 & \phantom{00}0.01 & \phantom{00}0.00 & \phantom{00}0.00 & \phantom{00}0.00 & \phantom{00}0.00 & \phantom{00}0.00 & \phantom{00}0.73 & \phantom{00}0.75 & \phantom{00}0.88 & \phantom{00}0.00 & \phantom{00}0.00 & \textbf{\phantom{00}0.91} & \phantom{00}0.83 & \textbf{\phantom{00}0.91} \\
Correct methods (e) & \phantom{00}0.85 & \phantom{00}0.88 & \phantom{00}0.81 & \phantom{00}0.93 & \phantom{00}0.87 & \phantom{00}0.83 & \phantom{00}0.88 & \phantom{00}0.93 & \phantom{00}0.87 & \phantom{00}0.84 & \textbf{1.00} & \phantom{00}0.00 & \phantom{00}0.00 & \phantom{00}0.00 & \phantom{00}0.00 & \phantom{00}0.00 & \phantom{00}0.74 & \phantom{00}0.84 & \phantom{00}0.89 & \phantom{00}0.00 & \phantom{00}0.00 & \phantom{00}0.92 & \phantom{00}0.83 & \phantom{00}0.91 \\
Illegal methods (t) & \phantom{00}0.12 & \phantom{00}0.09 & \phantom{00}0.12 & \phantom{00}0.05 & \phantom{00}0.11 & \phantom{00}0.12 & \phantom{00}0.10 & \phantom{00}0.03 & \phantom{00}0.10 & \phantom{00}0.13 & \textbf{\phantom{00}0.00} & \textbf{\phantom{00}0.00} & \textbf{\phantom{00}0.00} & \textbf{\phantom{00}0.00} & \textbf{\phantom{00}0.00} & \textbf{\phantom{00}0.00} & \phantom{00}0.22 & \phantom{00}0.13 & \phantom{00}0.10 & \textbf{\phantom{00}0.00} & \textbf{\phantom{00}0.00} & \phantom{00}0.07 & \phantom{00}0.14 & \phantom{00}0.07 \\
Illegal methods (e) & \phantom{00}0.13 & \phantom{00}0.09 & \phantom{00}0.12 & \phantom{00}0.05 & \phantom{00}0.11 & \phantom{00}0.12 & \phantom{00}0.11 & \phantom{00}0.05 & \phantom{00}0.10 & \phantom{00}0.14 & \textbf{\phantom{00}0.00} & \textbf{\phantom{00}0.00} & \textbf{\phantom{00}0.00} & \textbf{\phantom{00}0.00} & \textbf{\phantom{00}0.00} & \textbf{\phantom{00}0.00} & \phantom{00}0.23 & \phantom{00}0.14 & \phantom{00}0.10 & \textbf{\phantom{00}0.00} & \textbf{\phantom{00}0.00} & \phantom{00}0.07 & \phantom{00}0.14 & \phantom{00}0.07 \\
Correct argument names (t) & \phantom{00}0.39 & \phantom{00}0.50 & \phantom{00}0.49 & \phantom{00}0.41 & \phantom{00}0.63 & \phantom{00}0.45 & \phantom{00}0.53 & \phantom{00}0.25 & \phantom{00}0.58 & \phantom{00}0.52 & \phantom{00}0.00 & \phantom{00}0.00 & \phantom{00}0.00 & \phantom{00}0.00 & \phantom{00}0.00 & \phantom{00}0.00 & \phantom{00}0.37 & \phantom{00}0.56 & \phantom{00}0.68 & \phantom{00}0.00 & \phantom{00}0.00 & \phantom{00}0.79 & \phantom{00}0.76 & \textbf{\phantom{00}0.86} \\
Correct argument names (e) & \phantom{00}0.41 & \phantom{00}0.54 & \phantom{00}0.51 & \phantom{00}0.43 & \phantom{00}0.63 & \phantom{00}0.47 & \phantom{00}0.56 & \phantom{00}0.48 & \phantom{00}0.59 & \phantom{00}0.55 & \phantom{00}0.29 & \phantom{00}0.00 & \phantom{00}0.00 & \phantom{00}0.00 & \phantom{00}0.00 & \phantom{00}0.00 & \phantom{00}0.38 & \phantom{00}0.63 & \phantom{00}0.68 & \phantom{00}0.00 & \phantom{00}0.00 & \phantom{00}0.80 & \phantom{00}0.76 & \textbf{\phantom{00}0.87} \\
Correct argument values (t) & \phantom{00}0.29 & \phantom{00}0.43 & \phantom{00}0.40 & \phantom{00}0.33 & \phantom{00}0.58 & \phantom{00}0.39 & \phantom{00}0.47 & \phantom{00}0.18 & \phantom{00}0.53 & \phantom{00}0.48 & \phantom{00}0.00 & \phantom{00}0.00 & \phantom{00}0.00 & \phantom{00}0.00 & \phantom{00}0.00 & \phantom{00}0.00 & \phantom{00}0.33 & \phantom{00}0.39 & \phantom{00}0.63 & \phantom{00}0.00 & \phantom{00}0.00 & \phantom{00}0.76 & \phantom{00}0.72 & \textbf{\phantom{00}0.83} \\
Correct argument values (e) & \phantom{00}0.30 & \phantom{00}0.46 & \phantom{00}0.42 & \phantom{00}0.35 & \phantom{00}0.58 & \phantom{00}0.41 & \phantom{00}0.50 & \phantom{00}0.34 & \phantom{00}0.54 & \phantom{00}0.50 & \phantom{00}0.29 & \phantom{00}0.00 & \phantom{00}0.00 & \phantom{00}0.00 & \phantom{00}0.00 & \phantom{00}0.00 & \phantom{00}0.33 & \phantom{00}0.43 & \phantom{00}0.63 & \phantom{00}0.00 & \phantom{00}0.00 & \phantom{00}0.77 & \phantom{00}0.72 & \textbf{\phantom{00}0.83} \\
Missing arguments (t) & \phantom{00}0.61 & \phantom{00}0.50 & \phantom{00}0.51 & \phantom{00}0.59 & \phantom{00}0.37 & \phantom{00}0.55 & \phantom{00}0.47 & \phantom{00}0.75 & \phantom{00}0.42 & \phantom{00}0.48 & 1.00 & 1.00 & 1.00 & 1.00 & 1.00 & 1.00 & \phantom{00}0.63 & \phantom{00}0.44 & \phantom{00}0.32 & 1.00 & 1.00 & \phantom{00}0.21 & \phantom{00}0.24 & \textbf{\phantom{00}0.14} \\
Missing arguments (e) & \phantom{00}0.59 & \phantom{00}0.46 & \phantom{00}0.49 & \phantom{00}0.57 & \phantom{00}0.37 & \phantom{00}0.53 & \phantom{00}0.44 & \phantom{00}0.52 & \phantom{00}0.41 & \phantom{00}0.45 & \phantom{00}0.71 & \textbf{\phantom{00}0.00} & \textbf{\phantom{00}0.00} & \textbf{\phantom{00}0.00} & \textbf{\phantom{00}0.00} & \textbf{\phantom{00}0.00} & \phantom{00}0.62 & \phantom{00}0.37 & \phantom{00}0.32 & \textbf{\phantom{00}0.00} & \textbf{\phantom{00}0.00} & \phantom{00}0.20 & \phantom{00}0.24 & \phantom{00}0.13 \\
Unexpected arguments (t) & \phantom{00}0.30 & \phantom{00}0.25 & \phantom{00}0.34 & \phantom{00}0.31 & \phantom{00}0.25 & \phantom{00}0.32 & \phantom{00}0.25 & \phantom{00}0.11 & \phantom{00}0.24 & \phantom{00}0.22 & \textbf{\phantom{00}0.00} & \textbf{\phantom{00}0.00} & \textbf{\phantom{00}0.00} & \textbf{\phantom{00}0.00} & \textbf{\phantom{00}0.00} & \textbf{\phantom{00}0.00} & \phantom{00}0.17 & \phantom{00}0.22 & \phantom{00}0.20 & \textbf{\phantom{00}0.00} & \textbf{\phantom{00}0.00} & \phantom{00}0.18 & \phantom{00}0.17 & \phantom{00}0.11 \\
Unexpected arguments (e) & \phantom{00}0.31 & \phantom{00}0.26 & \phantom{00}0.35 & \phantom{00}0.32 & \phantom{00}0.25 & \phantom{00}0.33 & \phantom{00}0.26 & \phantom{00}0.20 & \phantom{00}0.25 & \phantom{00}0.23 & \phantom{00}0.12 & \textbf{\phantom{00}0.00} & \textbf{\phantom{00}0.00} & \textbf{\phantom{00}0.00} & \textbf{\phantom{00}0.00} & \textbf{\phantom{00}0.00} & \phantom{00}0.17 & \phantom{00}0.24 & \phantom{00}0.20 & \textbf{\phantom{00}0.00} & \textbf{\phantom{00}0.00} & \phantom{00}0.18 & \phantom{00}0.17 & \phantom{00}0.11 \\
Mean argument precision (t) & \phantom{00}0.50 & \phantom{00}0.61 & \phantom{00}0.53 & \phantom{00}0.49 & \phantom{00}0.67 & \phantom{00}0.53 & \phantom{00}0.62 & \phantom{00}0.36 & \phantom{00}0.59 & \phantom{00}0.63 & \phantom{00}0.01 & \phantom{00}0.00 & \phantom{00}0.00 & \phantom{00}0.00 & \phantom{00}0.00 & \phantom{00}0.00 & \phantom{00}0.56 & \phantom{00}0.65 & \phantom{00}0.75 & \phantom{00}0.00 & \phantom{00}0.00 & \phantom{00}0.82 & \phantom{00}0.82 & \textbf{\phantom{00}0.89} \\
Mean argument precision (e) & \phantom{00}0.52 & \phantom{00}0.65 & \phantom{00}0.54 & \phantom{00}0.51 & \phantom{00}0.68 & \phantom{00}0.55 & \phantom{00}0.65 & \phantom{00}0.63 & \phantom{00}0.60 & \phantom{00}0.67 & \phantom{00}0.67 & \phantom{00}0.00 & \phantom{00}0.00 & \phantom{00}0.00 & \phantom{00}0.00 & \phantom{00}0.00 & \phantom{00}0.57 & \phantom{00}0.72 & \phantom{00}0.75 & \phantom{00}0.00 & \phantom{00}0.00 & \phantom{00}0.82 & \phantom{00}0.82 & \textbf{\phantom{00}0.89} \\
Mean argument recall (t) & \phantom{00}0.41 & \phantom{00}0.51 & \phantom{00}0.48 & \phantom{00}0.42 & \phantom{00}0.63 & \phantom{00}0.47 & \phantom{00}0.54 & \phantom{00}0.28 & \phantom{00}0.56 & \phantom{00}0.55 & \phantom{00}0.00 & \phantom{00}0.00 & \phantom{00}0.00 & \phantom{00}0.00 & \phantom{00}0.00 & \phantom{00}0.00 & \phantom{00}0.40 & \phantom{00}0.58 & \phantom{00}0.69 & \phantom{00}0.00 & \phantom{00}0.00 & \phantom{00}0.81 & \phantom{00}0.78 & \textbf{\phantom{00}0.87} \\
Mean argument recall (e) & \phantom{00}0.43 & \phantom{00}0.54 & \phantom{00}0.50 & \phantom{00}0.44 & \phantom{00}0.63 & \phantom{00}0.49 & \phantom{00}0.56 & \phantom{00}0.49 & \phantom{00}0.56 & \phantom{00}0.57 & \phantom{00}0.33 & \phantom{00}0.00 & \phantom{00}0.00 & \phantom{00}0.00 & \phantom{00}0.00 & \phantom{00}0.00 & \phantom{00}0.41 & \phantom{00}0.65 & \phantom{00}0.69 & \phantom{00}0.00 & \phantom{00}0.00 & \phantom{00}0.82 & \phantom{00}0.78 & \textbf{\phantom{00}0.87} \\
Mean arg. Jaccard index (t) & \phantom{00}0.34 & \phantom{00}0.46 & \phantom{00}0.40 & \phantom{00}0.35 & \phantom{00}0.56 & \phantom{00}0.38 & \phantom{00}0.48 & \phantom{00}0.25 & \phantom{00}0.47 & \phantom{00}0.49 & \phantom{00}0.00 & \phantom{00}0.00 & \phantom{00}0.00 & \phantom{00}0.00 & \phantom{00}0.00 & \phantom{00}0.00 & \phantom{00}0.37 & \phantom{00}0.52 & \phantom{00}0.62 & \phantom{00}0.00 & \phantom{00}0.00 & \phantom{00}0.75 & \phantom{00}0.72 & \textbf{\phantom{00}0.83} \\
Mean arg. Jaccard index (e) & \phantom{00}0.36 & \phantom{00}0.48 & \phantom{00}0.41 & \phantom{00}0.37 & \phantom{00}0.56 & \phantom{00}0.40 & \phantom{00}0.50 & \phantom{00}0.44 & \phantom{00}0.48 & \phantom{00}0.52 & \phantom{00}0.33 & \phantom{00}0.00 & \phantom{00}0.00 & \phantom{00}0.00 & \phantom{00}0.00 & \phantom{00}0.00 & \phantom{00}0.37 & \phantom{00}0.58 & \phantom{00}0.63 & \phantom{00}0.00 & \phantom{00}0.00 & \phantom{00}0.75 & \phantom{00}0.72 & \textbf{\phantom{00}0.83} \\
Mean arg. val. cond. acc. (t) & \phantom{00}0.51 & \phantom{00}0.64 & \phantom{00}0.60 & \phantom{00}0.57 & \phantom{00}0.78 & \phantom{00}0.66 & \phantom{00}0.69 & \phantom{00}0.31 & \phantom{00}0.74 & \phantom{00}0.72 & \phantom{00}0.01 & \phantom{00}0.00 & \phantom{00}0.00 & \phantom{00}0.00 & \phantom{00}0.00 & \phantom{00}0.00 & \phantom{00}0.58 & \phantom{00}0.51 & \phantom{00}0.84 & \phantom{00}0.00 & \phantom{00}0.00 & \textbf{\phantom{00}0.95} & \textbf{\phantom{00}0.95} & \textbf{\phantom{00}0.95} \\
Mean arg. val. cond. acc. (e) & \phantom{00}0.53 & \phantom{00}0.68 & \phantom{00}0.62 & \phantom{00}0.60 & \phantom{00}0.79 & \phantom{00}0.69 & \phantom{00}0.72 & \phantom{00}0.54 & \phantom{00}0.75 & \phantom{00}0.76 & \phantom{00}0.67 & \phantom{00}0.00 & \phantom{00}0.00 & \phantom{00}0.00 & \phantom{00}0.00 & \phantom{00}0.00 & \phantom{00}0.59 & \phantom{00}0.57 & \phantom{00}0.84 & \phantom{00}0.00 & \phantom{00}0.00 & \textbf{\phantom{00}0.96} & \textbf{\phantom{00}0.96} & \phantom{00}0.95 \\
Total errors & 18\phantom{.00} & 22\phantom{.00} & 13\phantom{.00} & 21\phantom{.00} & 3\phantom{.00} & 16\phantom{.00} & 16\phantom{.00} & 169\phantom{.00} & 6\phantom{.00} & 18\phantom{.00} & 392\phantom{.00} & 395\phantom{.00} & 393\phantom{.00} & 395\phantom{.00} & 395\phantom{.00} & 390\phantom{.00} & 5\phantom{.00} & 42\phantom{.00} & 2\phantom{.00} & 395\phantom{.00} & 395\phantom{.00} & \textbf{1\phantom{.00}} & \textbf{1\phantom{.00}} & \textbf{1\phantom{.00}} \\
Incomplete implementations & 14\phantom{.00} & 5\phantom{.00} & 11\phantom{.00} & 5\phantom{.00} & 2\phantom{.00} & 12\phantom{.00} & 3\phantom{.00} & 2\phantom{.00} & 6\phantom{.00} & 9\phantom{.00} & 29\phantom{.00} & 35\phantom{.00} & 12\phantom{.00} & 2\phantom{.00} & 160\phantom{.00} & \textbf{0\phantom{.00}} & 4\phantom{.00} & 5\phantom{.00} & 1\phantom{.00} & 4\phantom{.00} & \textbf{0\phantom{.00}} & 1\phantom{.00} & \textbf{0\phantom{.00}} & \textbf{0\phantom{.00}} \\
Runtime errors & 4\phantom{.00} & 17\phantom{.00} & 2\phantom{.00} & 16\phantom{.00} & 1\phantom{.00} & 4\phantom{.00} & 13\phantom{.00} & 167\phantom{.00} & \textbf{0\phantom{.00}} & 9\phantom{.00} & 363\phantom{.00} & 360\phantom{.00} & 381\phantom{.00} & 393\phantom{.00} & 235\phantom{.00} & 390\phantom{.00} & 1\phantom{.00} & 37\phantom{.00} & 1\phantom{.00} & 391\phantom{.00} & 395\phantom{.00} & \textbf{0\phantom{.00}} & 1\phantom{.00} & 1\phantom{.00} \\
\bottomrule
\end{tabular}
}
}
\end{table}
\end{landscape}

\begin{landscape}
\begin{table}[p]
\centering
\caption{Complete Evaluation Results for \emph{Argument Completion}}
\label{tab:res-multi-ud-end}
\makebox[1\textwidth][c]{
\resizebox{\tablescale\textwidth}{!}{
\begin{tabular}{lrrrrrrrrrrrrrrrrrrrrrrrr}
\makebox[20pt][l]{\rotatebox{45}{}} & \makebox[20pt][l]{\rotatebox{45}{CodeT5+ (16B)}} & \makebox[20pt][l]{\rotatebox{45}{StarCoder (15.5B)}} & \makebox[20pt][l]{\rotatebox{45}{StarCoder2 (3B)}} & \makebox[20pt][l]{\rotatebox{45}{StarCoder2 (7B)}} & \makebox[20pt][l]{\rotatebox{45}{StarCoder2 (15B)}} & \makebox[20pt][l]{\rotatebox{45}{DeepSeek-Coder (1.3B)}} & \makebox[20pt][l]{\rotatebox{45}{DeepSeek-Coder (6.7B)}} & \makebox[20pt][l]{\rotatebox{45}{DeepSeek-Coder (7B)}} & \makebox[20pt][l]{\rotatebox{45}{DeepSeek-Coder (33B)}} & \makebox[20pt][l]{\rotatebox{45}{DeepSeek-Coder-V2 (16B)}} & \makebox[20pt][l]{\rotatebox{45}{Qwen2.5-Coder (0.5B)}} & \makebox[20pt][l]{\rotatebox{45}{Qwen2.5-Coder (1.5B)}} & \makebox[20pt][l]{\rotatebox{45}{Qwen2.5-Coder (3B)}} & \makebox[20pt][l]{\rotatebox{45}{Qwen2.5-Coder (7B)}} & \makebox[20pt][l]{\rotatebox{45}{Qwen2.5-Coder (14B)}} & \makebox[20pt][l]{\rotatebox{45}{Qwen2.5-Coder (32B)}} & \makebox[20pt][l]{\rotatebox{45}{Code Llama (7B)}} & \makebox[20pt][l]{\rotatebox{45}{Code Llama (13B)}} & \makebox[20pt][l]{\rotatebox{45}{Code Llama (70B)}} & \makebox[20pt][l]{\rotatebox{45}{Llama 3.1 (8B)}} & \makebox[20pt][l]{\rotatebox{45}{Llama 3.1 (70B)}} & \makebox[20pt][l]{\rotatebox{45}{Gemini 1.5 Pro}} & \makebox[20pt][l]{\rotatebox{45}{GPT-4o mini}} & \makebox[20pt][l]{\rotatebox{45}{GPT-4o}} \\
\midrule
Executable implementations (t) & \phantom{00}0.94 & \phantom{00}0.97 & \phantom{00}0.98 & \phantom{00}0.99 & \phantom{00}0.99 & \phantom{00}0.97 & \phantom{00}0.99 & \phantom{00}0.99 & \phantom{00}0.99 & \phantom{00}0.98 & \phantom{00}0.89 & \phantom{00}0.95 & \phantom{00}0.92 & \phantom{00}0.99 & \phantom{00}0.93 & \phantom{00}0.99 & \phantom{00}0.99 & \phantom{00}0.95 & \textbf{1.00} & \phantom{00}0.98 & \phantom{00}0.99 & \textbf{1.00} & \phantom{00}0.99 & \textbf{1.00} \\
Correct implementations (t) & \phantom{00}0.12 & \phantom{00}0.27 & \phantom{00}0.21 & \phantom{00}0.19 & \phantom{00}0.25 & \phantom{00}0.21 & \phantom{00}0.26 & \phantom{00}0.08 & \phantom{00}0.25 & \phantom{00}0.35 & \phantom{00}0.08 & \phantom{00}0.03 & \phantom{00}0.16 & \phantom{00}0.28 & \phantom{00}0.37 & \phantom{00}0.38 & \phantom{00}0.23 & \phantom{00}0.12 & \phantom{00}0.40 & \phantom{00}0.05 & \phantom{00}0.29 & \phantom{00}0.61 & \phantom{00}0.63 & \textbf{\phantom{00}0.77} \\
Correct implementations (e) & \phantom{00}0.13 & \phantom{00}0.27 & \phantom{00}0.22 & \phantom{00}0.19 & \phantom{00}0.25 & \phantom{00}0.22 & \phantom{00}0.27 & \phantom{00}0.08 & \phantom{00}0.25 & \phantom{00}0.36 & \phantom{00}0.09 & \phantom{00}0.03 & \phantom{00}0.18 & \phantom{00}0.28 & \phantom{00}0.40 & \phantom{00}0.38 & \phantom{00}0.23 & \phantom{00}0.13 & \phantom{00}0.40 & \phantom{00}0.05 & \phantom{00}0.29 & \phantom{00}0.61 & \phantom{00}0.64 & \textbf{\phantom{00}0.77} \\
Illegal implementations (t) & \phantom{00}0.59 & \phantom{00}0.27 & \phantom{00}0.48 & \phantom{00}0.47 & \phantom{00}0.35 & \phantom{00}0.51 & \phantom{00}0.33 & \phantom{00}0.21 & \phantom{00}0.38 & \phantom{00}0.31 & \phantom{00}0.58 & \phantom{00}0.50 & \phantom{00}0.44 & \phantom{00}0.41 & \phantom{00}0.32 & \phantom{00}0.32 & \phantom{00}0.11 & \phantom{00}0.28 & \phantom{00}0.23 & \phantom{00}0.30 & \phantom{00}0.26 & \phantom{00}0.17 & \phantom{00}0.20 & \textbf{\phantom{00}0.09} \\
Illegal implementations (e) & \phantom{00}0.63 & \phantom{00}0.27 & \phantom{00}0.49 & \phantom{00}0.47 & \phantom{00}0.36 & \phantom{00}0.52 & \phantom{00}0.33 & \phantom{00}0.22 & \phantom{00}0.39 & \phantom{00}0.32 & \phantom{00}0.65 & \phantom{00}0.53 & \phantom{00}0.48 & \phantom{00}0.41 & \phantom{00}0.35 & \phantom{00}0.32 & \phantom{00}0.11 & \phantom{00}0.30 & \phantom{00}0.23 & \phantom{00}0.31 & \phantom{00}0.26 & \phantom{00}0.17 & \phantom{00}0.20 & \textbf{\phantom{00}0.09} \\
Correct argument names (t) & \phantom{00}0.49 & \phantom{00}0.71 & \phantom{00}0.62 & \phantom{00}0.53 & \phantom{00}0.64 & \phantom{00}0.60 & \phantom{00}0.66 & \phantom{00}0.52 & \phantom{00}0.69 & \phantom{00}0.67 & \phantom{00}0.43 & \phantom{00}0.47 & \phantom{00}0.44 & \phantom{00}0.67 & \phantom{00}0.65 & \phantom{00}0.80 & \phantom{00}0.48 & \phantom{00}0.70 & \phantom{00}0.78 & \phantom{00}0.43 & \phantom{00}0.63 & \phantom{00}0.90 & \phantom{00}0.88 & \textbf{\phantom{00}0.93} \\
Correct argument names (e) & \phantom{00}0.52 & \phantom{00}0.73 & \phantom{00}0.63 & \phantom{00}0.54 & \phantom{00}0.65 & \phantom{00}0.62 & \phantom{00}0.67 & \phantom{00}0.53 & \phantom{00}0.69 & \phantom{00}0.69 & \phantom{00}0.49 & \phantom{00}0.50 & \phantom{00}0.48 & \phantom{00}0.67 & \phantom{00}0.70 & \phantom{00}0.81 & \phantom{00}0.48 & \phantom{00}0.73 & \phantom{00}0.79 & \phantom{00}0.44 & \phantom{00}0.64 & \phantom{00}0.90 & \phantom{00}0.89 & \textbf{\phantom{00}0.93} \\
Correct argument values (t) & \phantom{00}0.42 & \phantom{00}0.61 & \phantom{00}0.52 & \phantom{00}0.44 & \phantom{00}0.57 & \phantom{00}0.52 & \phantom{00}0.60 & \phantom{00}0.45 & \phantom{00}0.64 & \phantom{00}0.62 & \phantom{00}0.34 & \phantom{00}0.41 & \phantom{00}0.40 & \phantom{00}0.63 & \phantom{00}0.58 & \phantom{00}0.73 & \phantom{00}0.42 & \phantom{00}0.51 & \phantom{00}0.72 & \phantom{00}0.38 & \phantom{00}0.58 & \phantom{00}0.85 & \phantom{00}0.84 & \textbf{\phantom{00}0.90} \\
Correct argument values (e) & \phantom{00}0.45 & \phantom{00}0.62 & \phantom{00}0.53 & \phantom{00}0.44 & \phantom{00}0.58 & \phantom{00}0.54 & \phantom{00}0.61 & \phantom{00}0.46 & \phantom{00}0.65 & \phantom{00}0.63 & \phantom{00}0.40 & \phantom{00}0.43 & \phantom{00}0.43 & \phantom{00}0.63 & \phantom{00}0.63 & \phantom{00}0.74 & \phantom{00}0.42 & \phantom{00}0.53 & \phantom{00}0.72 & \phantom{00}0.39 & \phantom{00}0.58 & \phantom{00}0.85 & \phantom{00}0.84 & \textbf{\phantom{00}0.90} \\
Missing arguments (t) & \phantom{00}0.51 & \phantom{00}0.29 & \phantom{00}0.38 & \phantom{00}0.47 & \phantom{00}0.36 & \phantom{00}0.40 & \phantom{00}0.34 & \phantom{00}0.48 & \phantom{00}0.31 & \phantom{00}0.33 & \phantom{00}0.57 & \phantom{00}0.53 & \phantom{00}0.56 & \phantom{00}0.33 & \phantom{00}0.35 & \phantom{00}0.20 & \phantom{00}0.52 & \phantom{00}0.30 & \phantom{00}0.22 & \phantom{00}0.57 & \phantom{00}0.37 & \phantom{00}0.10 & \phantom{00}0.12 & \textbf{\phantom{00}0.07} \\
Missing arguments (e) & \phantom{00}0.48 & \phantom{00}0.27 & \phantom{00}0.37 & \phantom{00}0.46 & \phantom{00}0.35 & \phantom{00}0.38 & \phantom{00}0.33 & \phantom{00}0.47 & \phantom{00}0.31 & \phantom{00}0.31 & \phantom{00}0.51 & \phantom{00}0.50 & \phantom{00}0.52 & \phantom{00}0.33 & \phantom{00}0.30 & \phantom{00}0.19 & \phantom{00}0.52 & \phantom{00}0.27 & \phantom{00}0.21 & \phantom{00}0.56 & \phantom{00}0.36 & \phantom{00}0.10 & \phantom{00}0.11 & \textbf{\phantom{00}0.07} \\
Unnecessary arguments (t) & \phantom{00}0.03 & \phantom{00}0.05 & \phantom{00}0.08 & \phantom{00}0.07 & \phantom{00}0.07 & \phantom{00}0.05 & \phantom{00}0.07 & \phantom{00}0.06 & \phantom{00}0.07 & \phantom{00}0.03 & \phantom{00}0.03 & \phantom{00}0.03 & \phantom{00}0.03 & \phantom{00}0.06 & \phantom{00}0.03 & \phantom{00}0.07 & \phantom{00}0.05 & \phantom{00}0.05 & \phantom{00}0.05 & \phantom{00}0.03 & \phantom{00}0.04 & \phantom{00}0.04 & \phantom{00}0.02 & \textbf{\phantom{00}0.01} \\
Unnecessary arguments (e) & \phantom{00}0.03 & \phantom{00}0.05 & \phantom{00}0.08 & \phantom{00}0.07 & \phantom{00}0.07 & \phantom{00}0.05 & \phantom{00}0.07 & \phantom{00}0.06 & \phantom{00}0.07 & \phantom{00}0.03 & \phantom{00}0.03 & \phantom{00}0.03 & \phantom{00}0.03 & \phantom{00}0.06 & \phantom{00}0.03 & \phantom{00}0.07 & \phantom{00}0.05 & \phantom{00}0.05 & \phantom{00}0.05 & \phantom{00}0.03 & \phantom{00}0.04 & \phantom{00}0.04 & \phantom{00}0.02 & \textbf{\phantom{00}0.01} \\
Illegal arguments (t) & \phantom{00}0.24 & \phantom{00}0.13 & \phantom{00}0.21 & \phantom{00}0.20 & \phantom{00}0.14 & \phantom{00}0.23 & \phantom{00}0.15 & \phantom{00}0.08 & \phantom{00}0.13 & \phantom{00}0.13 & \phantom{00}0.29 & \phantom{00}0.24 & \phantom{00}0.23 & \phantom{00}0.17 & \phantom{00}0.18 & \phantom{00}0.11 & \phantom{00}0.06 & \phantom{00}0.13 & \phantom{00}0.09 & \phantom{00}0.16 & \phantom{00}0.12 & \phantom{00}0.09 & \phantom{00}0.09 & \textbf{\phantom{00}0.05} \\
Illegal arguments (e) & \phantom{00}0.25 & \phantom{00}0.13 & \phantom{00}0.21 & \phantom{00}0.20 & \phantom{00}0.14 & \phantom{00}0.24 & \phantom{00}0.15 & \phantom{00}0.09 & \phantom{00}0.13 & \phantom{00}0.13 & \phantom{00}0.31 & \phantom{00}0.25 & \phantom{00}0.24 & \phantom{00}0.17 & \phantom{00}0.20 & \phantom{00}0.11 & \phantom{00}0.06 & \phantom{00}0.13 & \phantom{00}0.09 & \phantom{00}0.16 & \phantom{00}0.12 & \phantom{00}0.09 & \phantom{00}0.09 & \textbf{\phantom{00}0.05} \\
Mean argument precision (t) & \phantom{00}0.55 & \phantom{00}0.79 & \phantom{00}0.60 & \phantom{00}0.61 & \phantom{00}0.71 & \phantom{00}0.64 & \phantom{00}0.73 & \phantom{00}0.73 & \phantom{00}0.69 & \phantom{00}0.76 & \phantom{00}0.48 & \phantom{00}0.59 & \phantom{00}0.58 & \phantom{00}0.71 & \phantom{00}0.69 & \phantom{00}0.78 & \phantom{00}0.66 & \phantom{00}0.75 & \phantom{00}0.83 & \phantom{00}0.63 & \phantom{00}0.78 & \phantom{00}0.88 & \phantom{00}0.89 & \textbf{\phantom{00}0.93} \\
Mean argument precision (e) & \phantom{00}0.58 & \phantom{00}0.81 & \phantom{00}0.62 & \phantom{00}0.61 & \phantom{00}0.71 & \phantom{00}0.65 & \phantom{00}0.74 & \phantom{00}0.74 & \phantom{00}0.69 & \phantom{00}0.77 & \phantom{00}0.54 & \phantom{00}0.62 & \phantom{00}0.63 & \phantom{00}0.71 & \phantom{00}0.74 & \phantom{00}0.78 & \phantom{00}0.66 & \phantom{00}0.79 & \phantom{00}0.83 & \phantom{00}0.65 & \phantom{00}0.79 & \phantom{00}0.88 & \phantom{00}0.90 & \textbf{\phantom{00}0.94} \\
Mean argument recall (t) & \phantom{00}0.50 & \phantom{00}0.72 & \phantom{00}0.61 & \phantom{00}0.53 & \phantom{00}0.64 & \phantom{00}0.63 & \phantom{00}0.66 & \phantom{00}0.52 & \phantom{00}0.66 & \phantom{00}0.70 & \phantom{00}0.46 & \phantom{00}0.46 & \phantom{00}0.46 & \phantom{00}0.68 & \phantom{00}0.67 & \phantom{00}0.78 & \phantom{00}0.53 & \phantom{00}0.70 & \phantom{00}0.78 & \phantom{00}0.42 & \phantom{00}0.66 & \phantom{00}0.90 & \phantom{00}0.90 & \textbf{\phantom{00}0.93} \\
Mean argument recall (e) & \phantom{00}0.53 & \phantom{00}0.74 & \phantom{00}0.62 & \phantom{00}0.54 & \phantom{00}0.65 & \phantom{00}0.64 & \phantom{00}0.67 & \phantom{00}0.52 & \phantom{00}0.67 & \phantom{00}0.71 & \phantom{00}0.52 & \phantom{00}0.49 & \phantom{00}0.50 & \phantom{00}0.68 & \phantom{00}0.72 & \phantom{00}0.79 & \phantom{00}0.53 & \phantom{00}0.74 & \phantom{00}0.79 & \phantom{00}0.43 & \phantom{00}0.67 & \phantom{00}0.90 & \phantom{00}0.90 & \textbf{\phantom{00}0.94} \\
Mean arg. Jaccard index (t) & \phantom{00}0.42 & \phantom{00}0.66 & \phantom{00}0.51 & \phantom{00}0.46 & \phantom{00}0.57 & \phantom{00}0.54 & \phantom{00}0.59 & \phantom{00}0.46 & \phantom{00}0.57 & \phantom{00}0.64 & \phantom{00}0.38 & \phantom{00}0.39 & \phantom{00}0.41 & \phantom{00}0.59 & \phantom{00}0.62 & \phantom{00}0.71 & \phantom{00}0.49 & \phantom{00}0.64 & \phantom{00}0.72 & \phantom{00}0.38 & \phantom{00}0.61 & \phantom{00}0.85 & \phantom{00}0.85 & \textbf{\phantom{00}0.91} \\
Mean arg. Jaccard index (e) & \phantom{00}0.45 & \phantom{00}0.68 & \phantom{00}0.52 & \phantom{00}0.47 & \phantom{00}0.57 & \phantom{00}0.55 & \phantom{00}0.60 & \phantom{00}0.47 & \phantom{00}0.57 & \phantom{00}0.65 & \phantom{00}0.43 & \phantom{00}0.41 & \phantom{00}0.45 & \phantom{00}0.59 & \phantom{00}0.67 & \phantom{00}0.71 & \phantom{00}0.49 & \phantom{00}0.67 & \phantom{00}0.73 & \phantom{00}0.39 & \phantom{00}0.61 & \phantom{00}0.85 & \phantom{00}0.86 & \textbf{\phantom{00}0.92} \\
Mean arg. val. cond. acc. (t) & \phantom{00}0.66 & \phantom{00}0.80 & \phantom{00}0.68 & \phantom{00}0.71 & \phantom{00}0.82 & \phantom{00}0.75 & \phantom{00}0.82 & \phantom{00}0.77 & \phantom{00}0.84 & \phantom{00}0.84 & \phantom{00}0.59 & \phantom{00}0.71 & \phantom{00}0.68 & \phantom{00}0.88 & \phantom{00}0.72 & \phantom{00}0.87 & \phantom{00}0.65 & \phantom{00}0.62 & \phantom{00}0.89 & \phantom{00}0.66 & \phantom{00}0.83 & \phantom{00}0.94 & \phantom{00}0.94 & \textbf{\phantom{00}0.95} \\
Mean arg. val. cond. acc. (e) & \phantom{00}0.71 & \phantom{00}0.83 & \phantom{00}0.70 & \phantom{00}0.72 & \phantom{00}0.83 & \phantom{00}0.77 & \phantom{00}0.83 & \phantom{00}0.78 & \phantom{00}0.84 & \phantom{00}0.86 & \phantom{00}0.67 & \phantom{00}0.74 & \phantom{00}0.74 & \phantom{00}0.89 & \phantom{00}0.78 & \phantom{00}0.87 & \phantom{00}0.65 & \phantom{00}0.65 & \phantom{00}0.89 & \phantom{00}0.67 & \phantom{00}0.84 & \phantom{00}0.94 & \textbf{\phantom{00}0.95} & \textbf{\phantom{00}0.95} \\
Total errors & 23\phantom{.00} & 11\phantom{.00} & 9\phantom{.00} & 4\phantom{.00} & 3\phantom{.00} & 11\phantom{.00} & 4\phantom{.00} & 5\phantom{.00} & 3\phantom{.00} & 6\phantom{.00} & 45\phantom{.00} & 19\phantom{.00} & 32\phantom{.00} & 2\phantom{.00} & 28\phantom{.00} & 2\phantom{.00} & 3\phantom{.00} & 19\phantom{.00} & \textbf{1\phantom{.00}} & 9\phantom{.00} & 5\phantom{.00} & \textbf{1\phantom{.00}} & 2\phantom{.00} & \textbf{1\phantom{.00}} \\
Incomplete implementations & 22\phantom{.00} & 5\phantom{.00} & 7\phantom{.00} & 4\phantom{.00} & 1\phantom{.00} & 11\phantom{.00} & 4\phantom{.00} & 3\phantom{.00} & 3\phantom{.00} & 6\phantom{.00} & 44\phantom{.00} & 17\phantom{.00} & 31\phantom{.00} & 2\phantom{.00} & 27\phantom{.00} & 2\phantom{.00} & 3\phantom{.00} & 5\phantom{.00} & 1\phantom{.00} & 6\phantom{.00} & 1\phantom{.00} & 1\phantom{.00} & \textbf{0\phantom{.00}} & \textbf{0\phantom{.00}} \\
Runtime errors & 1\phantom{.00} & 6\phantom{.00} & 2\phantom{.00} & \textbf{0\phantom{.00}} & 2\phantom{.00} & \textbf{0\phantom{.00}} & \textbf{0\phantom{.00}} & 2\phantom{.00} & \textbf{0\phantom{.00}} & \textbf{0\phantom{.00}} & 1\phantom{.00} & 2\phantom{.00} & 1\phantom{.00} & \textbf{0\phantom{.00}} & 1\phantom{.00} & \textbf{0\phantom{.00}} & \textbf{0\phantom{.00}} & 14\phantom{.00} & \textbf{0\phantom{.00}} & 3\phantom{.00} & 4\phantom{.00} & \textbf{0\phantom{.00}} & 2\phantom{.00} & 1\phantom{.00} \\
\bottomrule
\end{tabular}
}
}
\end{table}
\end{landscape}

\begin{table*}[p]
\centering
\caption{Evaluation Results of StarCoder2 by API for \emph{Full Completion}}
\label{tab:res-sc-ud-inv}
\begin{tabular}{lrrrrr}
\makebox[20pt][l]{\rotatebox{45}{}} & \makebox[20pt][l]{\rotatebox{45}{Overall}} & \makebox[20pt][l]{\rotatebox{45}{Asana}} & \makebox[20pt][l]{\rotatebox{45}{Google Calendar}} & \makebox[20pt][l]{\rotatebox{45}{Google Sheets}} & \makebox[20pt][l]{\rotatebox{45}{Slack}} \\
\midrule
Executable implementations (t) & 0.99 & 0.99 & \textbf{1.00} & \textbf{1.00} & 0.99 \\
Correct implementations (t) & 0.26 & 0.37 & \textbf{0.46} & 0.18 & 0.13 \\
Correct implementations (e) & 0.27 & 0.37 & \textbf{0.46} & 0.18 & 0.13 \\
Correct URLs (t) & 0.60 & 0.60 & \textbf{0.89} & 0.47 & 0.54 \\
Correct URLs (e) & 0.60 & 0.61 & \textbf{0.89} & 0.47 & 0.55 \\
Illegal URLs (t) & 0.22 & 0.25 & \textbf{0.03} & 0.18 & 0.22 \\
Illegal URLs (e) & 0.22 & 0.25 & \textbf{0.03} & 0.18 & 0.23 \\
Correct methods (t) & 0.86 & \textbf{0.90} & 0.89 & 0.47 & 0.86 \\
Correct methods (e) & 0.87 & \textbf{0.90} & 0.89 & 0.47 & 0.87 \\
Illegal methods (t) & 0.11 & 0.10 & \textbf{0.05} & 0.29 & 0.11 \\
Illegal methods (e) & 0.11 & 0.10 & \textbf{0.05} & 0.29 & 0.11 \\
Correct argument names (t) & 0.63 & 0.73 & \textbf{0.81} & 0.64 & 0.49 \\
Correct argument names (e) & 0.63 & 0.74 & \textbf{0.81} & 0.64 & 0.49 \\
Correct argument values (t) & 0.58 & 0.64 & \textbf{0.79} & 0.59 & 0.46 \\
Correct argument values (e) & 0.58 & 0.64 & \textbf{0.79} & 0.59 & 0.47 \\
Missing arguments (t) & 0.37 & 0.27 & \textbf{0.19} & 0.36 & 0.51 \\
Missing arguments (e) & 0.37 & 0.26 & \textbf{0.19} & 0.36 & 0.51 \\
Unexpected arguments (t) & 0.25 & \textbf{0.11} & 0.18 & 0.26 & 0.36 \\
Unexpected arguments (e) & 0.25 & \textbf{0.11} & 0.18 & 0.26 & 0.37 \\
Mean argument precision (t) & 0.67 & \textbf{0.88} & 0.76 & 0.60 & 0.46 \\
Mean argument precision (e) & 0.68 & \textbf{0.88} & 0.76 & 0.60 & 0.47 \\
Mean argument recall (t) & 0.63 & \textbf{0.76} & 0.75 & 0.66 & 0.47 \\
Mean argument recall (e) & 0.63 & \textbf{0.76} & 0.75 & 0.66 & 0.48 \\
Mean argument Jaccard index (t) & 0.56 & \textbf{0.71} & 0.68 & 0.52 & 0.38 \\
Mean argument Jaccard index (e) & 0.56 & \textbf{0.72} & 0.68 & 0.52 & 0.39 \\
Mean argument value conditional accuracy (t) & 0.78 & 0.85 & \textbf{0.87} & 0.84 & 0.70 \\
Mean argument value conditional accuracy (e) & 0.79 & 0.85 & \textbf{0.87} & 0.84 & 0.70 \\
Total errors & 3\phantom{.00} & 1\phantom{.00} & \textbf{0\phantom{.00}} & \textbf{0\phantom{.00}} & 2\phantom{.00} \\
Incomplete implementations & 2\phantom{.00} & 1\phantom{.00} & \textbf{0\phantom{.00}} & \textbf{0\phantom{.00}} & 1\phantom{.00} \\
Runtime errors & 1\phantom{.00} & \textbf{0\phantom{.00}} & \textbf{0\phantom{.00}} & \textbf{0\phantom{.00}} & 1\phantom{.00} \\
\bottomrule
\end{tabular}
\end{table*}

\begin{table*}[p]
\centering
\caption{Evaluation Results of StarCoder2 by API for \emph{Argument Completion}}
\label{tab:res-sc-ud-end}
\begin{tabular}{lrrrrr}
\makebox[20pt][l]{\rotatebox{45}{}} & \makebox[20pt][l]{\rotatebox{45}{Overall}} & \makebox[20pt][l]{\rotatebox{45}{Asana}} & \makebox[20pt][l]{\rotatebox{45}{Google Calendar}} & \makebox[20pt][l]{\rotatebox{45}{Google Sheets}} & \makebox[20pt][l]{\rotatebox{45}{Slack}} \\
\midrule
Executable implementations (t) & 0.99 & 0.99 & \textbf{1.00} & \textbf{1.00} & 0.99 \\
Correct implementations (t) & 0.25 & 0.37 & \textbf{0.46} & 0.24 & 0.10 \\
Correct implementations (e) & 0.25 & 0.37 & \textbf{0.46} & 0.24 & 0.10 \\
Illegal implementations (t) & 0.35 & 0.23 & \textbf{0.08} & 0.24 & 0.55 \\
Illegal implementations (e) & 0.36 & 0.23 & \textbf{0.08} & 0.24 & 0.55 \\
Correct argument names (t) & 0.64 & 0.69 & \textbf{0.86} & 0.67 & 0.55 \\
Correct argument names (e) & 0.65 & 0.69 & \textbf{0.86} & 0.67 & 0.55 \\
Correct argument values (t) & 0.57 & 0.59 & \textbf{0.81} & 0.61 & 0.49 \\
Correct argument values (e) & 0.58 & 0.59 & \textbf{0.81} & 0.61 & 0.50 \\
Missing arguments (t) & 0.36 & 0.31 & \textbf{0.14} & 0.33 & 0.45 \\
Missing arguments (e) & 0.35 & 0.31 & \textbf{0.14} & 0.33 & 0.45 \\
Unnecessary arguments (t) & 0.07 & \textbf{0.03} & 0.10 & 0.06 & 0.10 \\
Unnecessary arguments (e) & 0.07 & \textbf{0.03} & 0.10 & 0.06 & 0.10 \\
Illegal arguments (t) & 0.14 & 0.11 & \textbf{0.02} & 0.06 & 0.20 \\
Illegal arguments (e) & 0.14 & 0.11 & \textbf{0.02} & 0.06 & 0.20 \\
Mean argument precision (t) & 0.71 & \textbf{0.84} & 0.82 & 0.83 & 0.55 \\
Mean argument precision (e) & 0.71 & \textbf{0.85} & 0.82 & 0.83 & 0.55 \\
Mean argument recall (t) & 0.64 & 0.72 & \textbf{0.80} & 0.68 & 0.53 \\
Mean argument recall (e) & 0.65 & 0.72 & \textbf{0.80} & 0.68 & 0.54 \\
Mean argument Jaccard index (t) & 0.57 & 0.68 & \textbf{0.74} & 0.64 & 0.42 \\
Mean argument Jaccard index (e) & 0.57 & 0.69 & \textbf{0.74} & 0.64 & 0.43 \\
Mean argument value conditional accuracy (t) & 0.82 & 0.84 & \textbf{0.91} & 0.85 & 0.78 \\
Mean argument value conditional accuracy (e) & 0.83 & 0.84 & \textbf{0.91} & 0.85 & 0.79 \\
Total errors & 3\phantom{.00} & 1\phantom{.00} & \textbf{0\phantom{.00}} & \textbf{0\phantom{.00}} & 2\phantom{.00} \\
Incomplete implementations & 1\phantom{.00} & \textbf{0\phantom{.00}} & \textbf{0\phantom{.00}} & \textbf{0\phantom{.00}} & 1\phantom{.00} \\
Runtime errors & 2\phantom{.00} & 1\phantom{.00} & \textbf{0\phantom{.00}} & \textbf{0\phantom{.00}} & 1\phantom{.00} \\
\bottomrule
\end{tabular}
\end{table*}

\begin{table*}[p]
\centering
\caption{Evaluation Results of GPT-4o by API for \emph{Full Completion}}
\label{tab:res-4o-ud-inv}
\begin{tabular}{lrrrrr}
\makebox[20pt][l]{\rotatebox{45}{}} & \makebox[20pt][l]{\rotatebox{45}{Overall}} & \makebox[20pt][l]{\rotatebox{45}{Asana}} & \makebox[20pt][l]{\rotatebox{45}{Google Calendar}} & \makebox[20pt][l]{\rotatebox{45}{Google Sheets}} & \makebox[20pt][l]{\rotatebox{45}{Slack}} \\
\midrule
Executable implementations (t) & \textbf{1.00} & 0.99 & \textbf{1.00} & \textbf{1.00} & \textbf{1.00} \\
Correct implementations (t) & 0.60 & 0.56 & \textbf{0.89} & 0.59 & 0.57 \\
Correct implementations (e) & 0.60 & 0.57 & \textbf{0.89} & 0.59 & 0.57 \\
Correct URLs (t) & 0.78 & 0.75 & \textbf{1.00} & 0.76 & 0.75 \\
Correct URLs (e) & 0.78 & 0.76 & \textbf{1.00} & 0.76 & 0.75 \\
Illegal URLs (t) & 0.09 & 0.12 & \textbf{0.00} & \textbf{0.00} & 0.10 \\
Illegal URLs (e) & 0.09 & 0.12 & \textbf{0.00} & \textbf{0.00} & 0.10 \\
Correct methods (t) & 0.91 & 0.90 & 0.89 & 0.88 & \textbf{0.93} \\
Correct methods (e) & 0.91 & 0.90 & 0.89 & 0.88 & \textbf{0.93} \\
Illegal methods (t) & 0.07 & 0.10 & \textbf{0.00} & \textbf{0.00} & 0.07 \\
Illegal methods (e) & 0.07 & 0.10 & \textbf{0.00} & \textbf{0.00} & 0.07 \\
Correct argument names (t) & 0.86 & 0.83 & \textbf{1.00} & 0.88 & 0.86 \\
Correct argument names (e) & 0.87 & 0.84 & \textbf{1.00} & 0.88 & 0.86 \\
Correct argument values (t) & 0.83 & 0.79 & \textbf{1.00} & 0.81 & 0.83 \\
Correct argument values (e) & 0.83 & 0.80 & \textbf{1.00} & 0.81 & 0.83 \\
Missing arguments (t) & 0.14 & 0.17 & \textbf{0.00} & 0.12 & 0.14 \\
Missing arguments (e) & 0.13 & 0.16 & \textbf{0.00} & 0.12 & 0.14 \\
Unexpected arguments (t) & 0.11 & 0.11 & \textbf{0.00} & 0.06 & 0.13 \\
Unexpected arguments (e) & 0.11 & 0.11 & \textbf{0.00} & 0.06 & 0.13 \\
Mean argument precision (t) & 0.89 & 0.90 & \textbf{1.00} & 0.93 & 0.85 \\
Mean argument precision (e) & 0.89 & 0.90 & \textbf{1.00} & 0.93 & 0.85 \\
Mean argument recall (t) & 0.87 & 0.85 & \textbf{1.00} & 0.89 & 0.86 \\
Mean argument recall (e) & 0.87 & 0.86 & \textbf{1.00} & 0.89 & 0.86 \\
Mean argument Jaccard index (t) & 0.83 & 0.81 & \textbf{1.00} & 0.86 & 0.81 \\
Mean argument Jaccard index (e) & 0.83 & 0.81 & \textbf{1.00} & 0.86 & 0.81 \\
Mean argument value conditional accuracy (t) & 0.95 & 0.94 & \textbf{1.00} & 0.93 & 0.95 \\
Mean argument value conditional accuracy (e) & 0.95 & 0.95 & \textbf{1.00} & 0.93 & 0.95 \\
Total errors & 1\phantom{.00} & 1\phantom{.00} & \textbf{0\phantom{.00}} & \textbf{0\phantom{.00}} & \textbf{0\phantom{.00}} \\
Incomplete implementations & \textbf{0\phantom{.00}} & \textbf{0\phantom{.00}} & \textbf{0\phantom{.00}} & \textbf{0\phantom{.00}} & \textbf{0\phantom{.00}} \\
Runtime errors & 1\phantom{.00} & 1\phantom{.00} & \textbf{0\phantom{.00}} & \textbf{0\phantom{.00}} & \textbf{0\phantom{.00}} \\
\bottomrule
\end{tabular}
\end{table*}

\begin{table*}[p]
\centering
\caption{Evaluation Results of GPT-4o by API for \emph{Argument Completion}}
\label{tab:res-4o-ud-end}
\begin{tabular}{lrrrrr}
\makebox[20pt][l]{\rotatebox{45}{}} & \makebox[20pt][l]{\rotatebox{45}{Overall}} & \makebox[20pt][l]{\rotatebox{45}{Asana}} & \makebox[20pt][l]{\rotatebox{45}{Google Calendar}} & \makebox[20pt][l]{\rotatebox{45}{Google Sheets}} & \makebox[20pt][l]{\rotatebox{45}{Slack}} \\
\midrule
Executable implementations (t) & \textbf{1.00} & 0.99 & \textbf{1.00} & \textbf{1.00} & \textbf{1.00} \\
Correct implementations (t) & 0.77 & 0.77 & \textbf{0.86} & 0.71 & 0.75 \\
Correct implementations (e) & 0.77 & 0.77 & \textbf{0.86} & 0.71 & 0.75 \\
Illegal implementations (t) & 0.09 & 0.07 & 0.14 & \textbf{0.06} & 0.11 \\
Illegal implementations (e) & 0.09 & 0.07 & 0.14 & \textbf{0.06} & 0.11 \\
Correct argument names (t) & 0.93 & \textbf{0.96} & 0.93 & 0.92 & 0.91 \\
Correct argument names (e) & 0.93 & \textbf{0.97} & 0.93 & 0.92 & 0.91 \\
Correct argument values (t) & 0.90 & 0.91 & \textbf{0.93} & 0.89 & 0.88 \\
Correct argument values (e) & 0.90 & 0.92 & \textbf{0.93} & 0.89 & 0.88 \\
Missing arguments (t) & 0.07 & \textbf{0.04} & 0.07 & 0.08 & 0.09 \\
Missing arguments (e) & 0.07 & \textbf{0.03} & 0.07 & 0.08 & 0.09 \\
Unnecessary arguments (t) & 0.01 & 0.02 & \textbf{0.00} & \textbf{0.00} & 0.02 \\
Unnecessary arguments (e) & 0.01 & 0.02 & \textbf{0.00} & \textbf{0.00} & 0.02 \\
Illegal arguments (t) & 0.05 & \textbf{0.02} & 0.07 & \textbf{0.02} & 0.07 \\
Illegal arguments (e) & 0.05 & \textbf{0.02} & 0.07 & \textbf{0.02} & 0.07 \\
Mean argument precision (t) & 0.93 & 0.96 & 0.93 & \textbf{0.98} & 0.90 \\
Mean argument precision (e) & 0.94 & 0.97 & 0.93 & \textbf{0.98} & 0.90 \\
Mean argument recall (t) & 0.93 & \textbf{0.97} & 0.93 & 0.93 & 0.91 \\
Mean argument recall (e) & 0.94 & \textbf{0.97} & 0.93 & 0.93 & 0.91 \\
Mean argument Jaccard index (t) & 0.91 & \textbf{0.95} & 0.91 & 0.92 & 0.88 \\
Mean argument Jaccard index (e) & 0.92 & \textbf{0.95} & 0.91 & 0.92 & 0.88 \\
Mean argument value conditional accuracy (t) & 0.95 & 0.95 & \textbf{1.00} & 0.96 & 0.94 \\
Mean argument value conditional accuracy (e) & 0.95 & 0.96 & \textbf{1.00} & 0.96 & 0.94 \\
Total errors & 1\phantom{.00} & 1\phantom{.00} & \textbf{0\phantom{.00}} & \textbf{0\phantom{.00}} & \textbf{0\phantom{.00}} \\
Incomplete implementations & \textbf{0\phantom{.00}} & \textbf{0\phantom{.00}} & \textbf{0\phantom{.00}} & \textbf{0\phantom{.00}} & \textbf{0\phantom{.00}} \\
Runtime errors & 1\phantom{.00} & 1\phantom{.00} & \textbf{0\phantom{.00}} & \textbf{0\phantom{.00}} & \textbf{0\phantom{.00}} \\
\bottomrule
\end{tabular}
\end{table*}

\end{document}